\documentclass[twocolumn,showpacs,preprintnumbers,prb,amsmath,amssymb]{revtex4}%PRB %\documentclass[preprint,showpacs,preprintnumbers,prl,amsmath,amssymb]{revtex4}%PRL
\usepackage{graphicx}% Include figure files
\usepackage{dcolumn}% Align table columns on decimal point
\usepackage{bm}% bold math
\usepackage{longtable}% tables which need more space than one page
\usepackage{array}
\usepackage{color}
\begin{document}

% \preprint{}
\title{Hole spin mixing in InAs Quantum Dot Molecules}
\author{M. F. Doty$^{1,2}$}
\email{doty@udel.edu}
\author{J. I. Climente$^{3}$}
\author{A. Greilich$^{2}$}
\author{A. S. Bracker$^{2}$}
\author{D. Gammon$^{2}$}
\affiliation{$^{1}$Dept. of Materials Science and Engineering, University of Delaware, Newark, DE 19716}
\affiliation{$^{2}$Naval Research Laboratory, Washington, DC 20375}
\affiliation{$^{3}$Dept. de Qu\'imica F\'isica i Anal\'itica, Universitat Jaume I, 12080, Castell\'o, Spain}
\date{\today}% It is always \today, today,
             %  but any date may be explicitly specified

\begin{abstract}
 Holes confined in single InAs quantum dots have recently emerged as a promising system for the storage or manipulation of quantum information. These holes are often assumed to have only heavy-hole character and further assumed to have no mixing between orthogonal heavy hole spin projections (in the absence of a transverse magnetic field). The same assumption has been applied to InAs quantum dot molecules formed by two stacked InAs quantum dots that are coupled by coherent tunneling of the hole between the two dots. We present experimental evidence of the existence of a hole spin mixing term obtained with magneto-photoluminescence spectroscopy on such InAs quantum dot molecules. We use a Luttinger spinor model to explain the physical origin of this hole spin mixing term: misalignment of the dots along the stacking direction breaks the angular symmetry and allows mixing through the light-hole component of the spinor. We discuss how this novel spin mixing mechanism may offer new spin manipulation opportunities that are unique to holes.
\end{abstract}

% PACS, the Physics and Astronomy
                             % Classification Scheme.
%\keywords{Suggested keywords}%Use showkeys class option if keyword
                              %display desired
\maketitle
% ***************************************************************
% *  Begin Main Text                                            *
% ***************************************************************
\section{Introduction}
\label{Sect:Intro}
An electron excited across the band gap of a semiconductor quantum
dot (QD) leaves behind a hole in the otherwise full valence states.  This
hole behaves like a charged particle, much like the electron, though
with a substantially larger effective mass. Because the valence states are derived from p-type atomic states of
the lattice, a hole experiences a strong spin-orbit (SO) interaction that leads to a new spin basis in which the low energy hole state has total angular momentum $J = 3/2$. Because light hole (LH) states ($J_z = \pm 1/2$) are shifted to larger energy by confinement and strain, it is generally a good approximation to treat low-energy holes in QDs as if they have only heavy-hole (HH) ($J_z = \pm 3/2$) character with a pseudo-spin $\frac{1}{2}$. This simple picture has proven to be remarkably useful and explains a wide variety of optical and magneto-optical properties of quantum dots.\cite{BayerPRB00, HeissPRB07} In the absence of a transverse magnetic field, the orthogonal HH spin projections do not mix. The well isolated spin projections and lack of a significant contact hyperfine interaction with nuclear spins makes the HH spin projection a good candidate for the storage of quantum information.\cite{GerardotNature08, BrunnerScience09}

Here we present experimental evidence of mixing between HH spin projections in coupled quantum dots. The mixing is observed in the optical spectra of stacked pairs of self-assembled InAs quantum dots near the applied electric field that induces coherent hole tunnelling between the QDs. Spin mixing between bright and dark exciton spin configurations causes dark states to gain optical intensity, which we have observed in a number of cases. Here we present an example in which the spin mixing is sufficiently large that we can directly observe anticrossings between bright and dark exciton states and measure the mixing between opposite heavy-hole spin projections. In Sect.~\ref{Sect:EnergyLevels} we present the energy levels for the neutral exciton states of a coupled pair of QDs and describe the expected bright and dark state energy levels and interactions in the absence of hole spin mixing. In Sect.~\ref{Sect:Evidence} we present experimental evidence of the appearance of bright-dark anticrossings and show that the experimental data can be phenomenologically explained by the presence of hole spin mixing. In Sect.~\ref{origin} we use a Luttinger spinor model to explain how the spin mixing can arise when misalignment of the QDs along the stacking axis breaks the molecular symmetry. A complete description of the theoretical model can be found in App.~\ref{app:theo}. In Sect.~\ref{conclusion} we summarize our observation and explanation of the origin of hole spin mixing.

To build a quantum information processing device around hole spins, it is necessary to have optical or electrical mechanisms for coherently creating and manipulating superpositions of orthogonal spin projections.\cite{EconomouPRL07, AndlauerPRB09, DotyPhys09} The novel spin-mixing mechanism presented here presents new opportunities for electrical control of hole spin projections, which are discussed in Sect.~\ref{Sect:Implications}. We calculate the purity of the hole spin states as a function of applied electric field and show that the heavy-hole states remain good states for the storage of quantum information away from the electric field of tunnel coupling. We suggest a possible scenario for electrically gating the hole spin mixing and thus manipulating the hole spin projection.  We note that an electron spin-mixing anticrossing was previously measured and used for optical spin control.\cite{KimPRL08} The hole spin anticrossing energy measured here is an order of magnitude larger.

\section{Energy level structure of the neutral exciton}
\label{Sect:EnergyLevels}

\begin{figure}[htb]
\includegraphics[width=8.6cm]{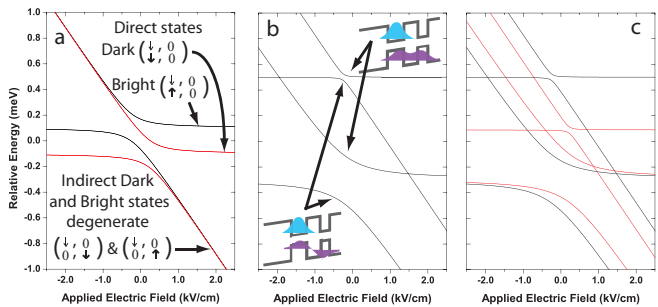}
\caption{(Color Online) a) Energy levels of bright and dark exciton states in a QDM at B = 0 T. b,c) Energy levels of bright states (b) and bright and dark states (c) at B = 6 T. Black (red) lines indicate optically bright (dark) exciton states.}
\label{ExcitonEnergy}
\end{figure}

When two InAs QDs are stacked on top of one another, electrons or holes can tunnel between the two dots to create quantum dot molecules (QDMs).\cite{BrackerAPL06} In general the tunnel coupling is weak because the natural distribution of dot size, shape, and alloying leads to different confined energy levels in each dot. An electric field applied along the growth direction can tune the energy levels into resonance to enable coherent tunneling of electrons or holes between the dots and the formation of delocalized states with molecular orbital character.\cite{KrennerPRL05} The experimental signature of the delocalized molecular orbitals is the formation of an anticrossing between photoluminescence (PL) lines that come from direct (electron and hole in the same quantum dot) and indirect (electron and hole in different quantum dots) states. Fig.~\ref{ExcitonEnergy}a shows a calculation of the energy levels and anticrossings for the neutral exciton ($X^0$: one electron and one hole) in a QDM where the dots are separated by a 4 nm barrier and the hole tunnels between dots. Direct and indirect states are labeled.

To denote the spatial locations of the electrons and holes we use the notation $\left( ^{e_B e_T}_{h_b h_T}\right)$, where $e_B$ ($e_T$) are the spins of the electrons in the bottom (top) dot and similarly for holes. This notation describes the sates far away from an anticrossing. At the anticrossings, the molecular states can be described as symmetric and antisymmetric combinations of these basis states. If we make the usual assumption that holes have only HH character ($J_z = \pm 3/2$), there are four possible pairings of the electron and hole spin projections in a neutral exciton. We use $\uparrow$, $\downarrow$ to indicate the electron spin projection ($S_e = \pm1/2$) and $\Uparrow$, $\Downarrow$ to indicate the HH spin projection ($J_z = \pm3/2$). The two spin pairings  $\uparrow\Downarrow$ and $\downarrow\Uparrow$ have exciton angular momentum $\pm1$ and can couple to photons. These exciton spin configurations are called bright states. The other two spin configurations ($\uparrow\Uparrow$ and $\downarrow\Downarrow$) have exciton angular momentum $\pm2$ and are called dark excitons because they do not couple to photons. One set of bright and dark states are labeled in Fig.~\ref{ExcitonEnergy}a using our notation. The direct bright and dark states are split due to the isotropic electron-hole exchange interaction. This interaction is suppressed when electrons and holes are in separate dots and thus the indirect bright and dark states are degenerate. In a typical experimental spectra, only the bright states are evident.

In the absence of a magnetic field or hole spin mixing, orthogonal electron and hole spin projections are degenerate. In this work we ignore the small mixing between the two bright states that arises from the anisotropic exchange interaction.\cite{GammonPRL96} The two bright exciton spin configurations are thus degenerate and only the electron spin down configuration is labeled in Fig.~\ref{ExcitonEnergy}a. One anticrossing is observed where  the direct and indirect bright exciton spin configurations would be expected to intersect (the upper anticrossing in Fig.~\ref{ExcitonEnergy}a). Another anticrossing is observed for the dark exciton states (lower anticrossing). Tunnel coupling, anticrossings, and spin interactions of electrons and holes in both neutral and charged excitons have been observed. \cite{StinaffScience06, KrennerPRL06, ScheibnerPRB07, DotyPRB08} In this work we focus on the tunnel coupling of holes in the neutral exciton state, as depicted in Fig.~\ref{ExcitonEnergy}a.

When a longitudinal magnetic field is applied (Faraday geometry: parallel to the optical axis), the natural expectation is that a Zeeman splitting between the two bright exciton configurations will be observed, with a splitting proportional to the sum of the electron and heavy hole g factors. However, in QDMs where holes tunnel the formation of molecular orbitals substantially alters the hole g factor.\cite{DotyPRL06, DotyPRL09, AndlauerPRB09} Consequently, the hole g factor depends on the applied electric field and strong resonant enhancement or suppression of the Zeeman splitting is observed at the electric field of coupling.\cite{DotyPRL06}

In Fig.~\ref{ExcitonEnergy}b we show the calculated energies of the bright states from Fig.~\ref{ExcitonEnergy}a when a magnetic field of 6 T is applied. The first effect of the magnetic field is a Zeeman splitting of the two bright states that were degenerate in Fig.~\ref{ExcitonEnergy}a. This is seen most clearly at the edges of Fig.~\ref{ExcitonEnergy}b: the degenerate lines at +0.1 meV in Fig.~\ref{ExcitonEnergy}a split and move to -0.3 meV and +0.5 meV. The second effect of the magnetic field is the introduction of a g factor resonance. The g factor resonance is seen most clearly by looking at the anticrossings. The degenerate anticrossings in Fig.~\ref{ExcitonEnergy}a have an anticrossing gap of 214 $\mu$eV. In Fig.~\ref{ExcitonEnergy}b, the anticrossing gap for the lower Zeeman branch expands to approximately 400 $\mu$eV, while the anticrossing of the upper Zeeman branch collapses to approximately 30 $\mu$eV.

The difference in anticrossing energies arises because different molecular orbitals have different g factor contributions from the barrier, which determine the net hole g factor.\cite{DotyPRL06, DotyPRL09} The two lower energy molecular states (one for each bright electron-hole spin orientation) have antisymmetric (noded) orbital character (lower inset to Fig.~\ref{ExcitonEnergy}b). The node in the molecular wavefunction suppresses the contribution of the GaAs barrier to the net hole g factor. Because the barrier hole g factor is positive, suppression of this contribution increases the relative weight of the negative g factor contribution from the InAs QDs. The increase in the magnitude of the g factor (more negative) causes the Zeeman splitting of the two antisymmetric (noded) molecular orbitals to increase on resonance. Conversely, the higher energy molecular states have symmetric (node-less) orbital character (upper inset). The contribution of the barrier is thus enhanced on resonance, offsetting the negative contribution from the InAs QDs and reducing the Zeeman splitting of these two lines. The combination of the enhanced Zeeman splitting for one molecular branch and suppressed Zeeman splitting for the other branch leads to the different anticrossing energies. In Fig.~\ref{ExcitonEnergy}c we plot the energies of both the bright and dark exciton states including the Zeeman splitting and the resonant changes in g factor. In the absence of hole spin mixing there are no anticrossings where bright and dark states intersect.

The counterintuitive antisymmetric (noded) character of the molecular ground states in Fig.~\ref{ExcitonEnergy}b is a consequence of the spin-orbit interaction, which mixes HH and LH states.\cite{DotyPRL09} When the barrier separating the QDs is thin, the molecular ground state has bonding (symmetric) orbital character and the first molecular excited state has antibonding (antisymmetric) character, in analogy with natural diatomic molecules. As the thickness of the barrier is increased (to 4 nm in Fig.~\ref{ExcitonEnergy}), the contribution of the LH states becomes more important and leads to the reversal of the orbital character. The reversal provides one indication that LHs can not be neglected in QDMs. To include LHs, hole states are described as Luttinger spinors that contain all four projections of $J_z$, but each spinor is dominated by a single HH spin projection.\cite{ClimentePRB08} As we describe below, the mixing that leads to the reversal of symmetric and anti-symmetric orbital states does not result in mixing between these HH spin projections and consequently cannot explain the appearance of bright-dark anticrossings. However, if misalignment of the dots along the stacking axis breaks the symmetry of the QDM, at electric fields near the point of tunnel coupling the spin-orbit interaction combines with the broken symmetry to permit mixing of spinors with different HH spin components.

\begin{figure}[htb]
\includegraphics[width=8.6cm]{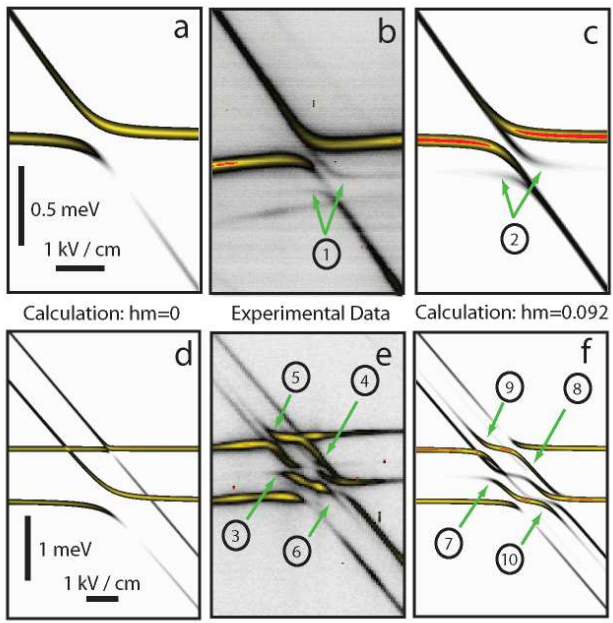}
\caption{(Color Online) Experimental (b) and calculated (a,c) photoluminescence spectral map of the neutral exciton at B = 0 T without (a) and with (c) phenomenological hole spin mixing term. (d-f) Spectral maps as in (a-c) with B = 6 T. Scales for panels (a-c) are the same and are indicated in panel a, similarly for (d-f). Callout numbers are referenced in the text.}
\label{ExptHoleMix}
\end{figure}

Fig.\ref{ExptHoleMix}a and d display calculated spectra of the anticrossings of the $X^0$ state at B = 0 and 6 T. These calculations use the same parameters as Fig.~\ref{ExcitonEnergy}, but include a color mapping to display the optical intensity of the lines. The calculated spectra in Fig.\ref{ExptHoleMix}a and d are representative of the typical experimentally observed spectra. To calculate these spectra we use matrix Hamiltonians, which have been shown to provide an accurate phenomenological model of tunneling, spin interactions, and resonant changes to g factor in a wide variety of samples.\cite{StinaffScience06, DotyPRL06, ScheibnerPRB07, ScheibnerPRL07, DotyPRB08, ScheibnerNatPhys08, DotyPRL09} The basis states are the possible spatial and spin distributions of the electron and hole. When an electric field is applied, the lowest electron energy level in the top dot is at significantly higher energy than the confined electron energy level of the bottom dot and the states with the electron in the top dot can be neglected. We can describe the basis states in our notation ($\left( ^{e_B e_T}_{h_b h_T}\right)$) with holes denoted by the dominant HH spin projection. Using this notation, the basis states of the Hamiltonian are:

\begin{equation}
\begin{array}{cccc}
\left(\begin{array}{cc}
\downarrow & 0\\
\Uparrow & 0\\
\end{array}\right)

&

\left(\begin{array}{cc}
\downarrow & 0\\
0 & \Uparrow\\
\end{array}\right)

&

\left(\begin{array}{cc}
\downarrow & 0\\
\Downarrow & 0\\
\end{array}\right)

&

\left(\begin{array}{cc}
\downarrow & 0\\
0 & \Downarrow \\
\end{array}\right)
\end{array}
\label{atomic_basis}
\end{equation}

Note that only the electron spin down case is shown because the overall matrix is block diagonal for the two electron spin projections. The first two states are bright excitons, the second two dark excitons. The Hamiltonian that describes the energy of the neutral exciton state is:
\begin{widetext}
\hspace{-1cm}\begin{equation}\begin{array}{cccc}
+\delta_0+\frac{\mu_B B (g_e+g_{hB})}{2} & -t_{X^0}+\frac{\mu_B B g_{12}}{2} & 0 & hm\\
 & & &\\
-t_{X^0}+\frac{\mu_B B g_{12}}{2} & -d F+ \frac{\mu_B B (g_e+g_{hT})}{2} & hm & 0\\
 & & &\\
 0 & hm & -\delta_0+\frac{\mu_B B (g_e-g_{hB})}{2} & -t_{X^0}-\frac{\mu_B B g_{12}}{2}\\
 & & &\\
hm & 0 & -t_{X^0}-\frac{\mu_B B g_{12}}{2} & -d F+ \frac{\mu_B B (g_e-g_{hT})}{2} \\
\end{array}\label{MatrixHamiltonian}
\end{equation}
%Note full matrix equation is written out in comments at end of document
\end{widetext}

$\delta_0$ is the electron-hole exchange interaction that splits bright and dark states when the electron and hole are in the same dot. $\mu_B$ is the Bohr magneton, B is the magnetic field. $g_e$ is the electron g factor in the bottom dot. $g_{hB}$ ($g_{hT}$) is the g factor for a hole in the bottom (top) dot. $t_{X^0}$ is the tunneling matrix element. $d$ is the effective barrier thickness, which determines the slope of the indirect lines when the electric field, $F$, is applied. $hm$ is the hole mixing term, which is set to zero for the calculations in Fig.\ref{ExptHoleMix}a and d. $g_{12}$ is the resonant contribution to the g factor from the barrier, which has opposite sign to $g_h$.\cite{DotyPRL09} The relative signs of $t_{X^0}$ and $g_{12}$ insure that the bonding orbital has a decreased splitting. Here $t_{X^0}<0$ because the molecular ground state has antibonding character.

The energies of the neutral exciton states are calculated by finding the
eigenvalues of the matrix at a specific value of the field, $F$. Because the final state after optical recombination contains no particles, the energies of the neutral exciton initial states are exactly the energies of the observed PL lines. The optical intensities are calculated by multiplying the corresponding eigenvector by an optical intensity vector:

\begin{equation}
\left(\begin{array}{c}
1\\
Ind/\sqrt{2}\\
0\\
0\\
\end{array}\right)
\end{equation}

which simply gives unit intensity to the direct transition (electron
and hole in the same dot) and a fraction of that intensity (\textit{Ind}) to
the indirect transition (electron and hole in different dots). The
dark states have no optical intensity. In the calculations, therefore, any optical intensity for dark states must come from mixing with bright states.

\section{Experimental evidence of hole spin mixing}
\label{Sect:Evidence}
The calculated PL spectral maps in Fig.\ref{ExptHoleMix}a and d are representative of the observed behavior for the anticrossing of the neutral exciton in most samples. As mentioned in Sect.~\ref{Sect:Intro}, we have observed a number of QDMs where the dark states gain optical intensity in the vicinity of the anticrossing region. This intensity gain cannot be explained with any of the previously observed QDM properties, including spin-conserving tunneling and molecular g factor resonances. Fig.\ref{ExptHoleMix}b and e shows experimental data for a QDM where the dots are separated by a 4 nm tunnel barrier. In this example, not only do the dark states gain optical intensity, we also observe new anticrossings between bright and dark states at high magnetic field. These anticrossings allow us to directly measure the magnitude of the spin mixing and to show that it is responsible for the dark state intensity. In the zero magnetic field case (Fig.\ref{ExptHoleMix}b) the new anticrossing with the dark exciton state is indicated by callout 1. At B = 6 T (Fig.\ref{ExptHoleMix}e) a complex pattern of additional anticrossings (callouts 3-6) appear near the electric field of the tunnel resonance. We will show that all of the additional anticrossings can be explained by the inclusion of the hole spin mixing term, which couples bright and dark exciton states.

The calculations in Fig.\ref{ExptHoleMix}a and d are obtained with numerical values determined by the experimental data in Fig.\ref{ExptHoleMix}b and e with the hole spin mixing term ($hm$) set to zero (see Appendix \ref{app:params}). It is clear that when the hole spin mixing term is set to zero the phenomenological Hamiltonian (Fig.\ref{ExptHoleMix}a and d) does not capture all of the features that appear in the experimental data. However, as shown in Fig.\ref{ExptHoleMix}c and f, all of the dark states and anticrossings in the experimental spectra are explained if we turn on the hole mixing term. Best agreement with the data is obtained when $hm = 92 \mu$eV, $t_{X^0}=0$ and all other parameters have the same value as in Fig.\ref{ExptHoleMix}a and d. The suppression of $t_{X^0}=0$ is discussed below.

Without the hole spin mixing term, bright and dark exciton configurations are independent. In this sample the intradot (direct) dark exciton configurations typically lie about 200 $\mu$eV below the bright exciton configurations.\footnote{For the particular QDM shown in Fig.~\ref{ExptHoleMix} the dark exciton lies 202 $\mu$eV below the bright exciton.} The anticrossing observed in Fig.\ref{ExptHoleMix}a occurs between the direct and indirect bright exciton states. Spin-conserving tunneling also couples the direct and indirect dark exciton states, but no signature appears in the PL spectra because the dark states do not couple to photons. The hole spin mixing term allows the spin-up hole in the bottom dot to mix with the spin-down hole in the top dot. This coupling mixes bright (e.g. $\downarrow\Uparrow$) and dark (e.g. $\downarrow\Downarrow$) exciton states and creates anticrossings wherever bright \textbf{or dark} exciton states would cross. The eigenstates with mostly dark exciton character gain optical intensity near the anticrossings because they contain nonzero bright exciton components as a result of the mixing.

In Fig.\ref{ExptHoleMix}b callout 1 points out the dark exciton states that have gained optical intensity in the experimental spectra. In Fig.\ref{ExptHoleMix}c callout 2 indicates that the inclusion of a the hole spin mixing term in the phenomenological Hamiltonian leads to the dark exciton states gaining optical intensity. The direct dark exciton state ($\left(^{\downarrow, \hspace{1pt}0}_{\Downarrow, \hspace{1pt}0}\right)$) is no longer an eigenstate of the system. As a result of the hole spin mixing, the new eigenstate includes a nonzero contribution from the indirect bright exciton state ($\left(^{\downarrow, \hspace{1pt}0}_{\hspace{1pt}0, \Uparrow}\right)$). It is this bright component that gives the eigenstate optical intensity. The appearance of the dark state at zero magnetic field can not be explained without the hole spin mixing term.

When a magnetic field is applied, both bright and dark exciton configurations undergo a Zeeman splitting. In the absence of hole spin mixing the bright and dark states simply cross and the dark states remain dark (Fig.\ref{ExptHoleMix}d). In the presence of hole spin mixing, each crossing of bright and dark states becomes an anticrossing observable in the experimental spectra. Callout 3 in Fig.~\ref{ExptHoleMix}e, for example, is an anticrossing between a direct dark and indirect bright exciton state. We can directly measure the magnitude of this anticrossing gap (180 $\mu$eV) to determine the magnitude of the spin mixing term. Fig.~\ref{ExptHoleMix}f shows that the inclusion of the spin mixing term explains the additional anticrossings observed in the experimental spectra. The anticrossings indicated by callouts 7 and 8 arise from the anticrossing of the indirect bright ($\left(^{\downarrow, \hspace{1pt}0}_{\hspace{1pt}0, \Uparrow}\right)$) and direct dark ($\left(^{\downarrow, \hspace{1pt}0}_{\Downarrow, \hspace{1pt}0}\right)$) excitons. These calculated anticrossings correspond to the observed anticrossings indicated by callouts 3 and 4 in Fig.~\ref{ExptHoleMix}e. The anticrossings indicated by callouts 9 and 10 arise from anticrossings between the direct bright ($\left(^{\downarrow, \hspace{1pt}0}_{\Uparrow, \hspace{1pt}0}\right)$) and indirect dark ($\left(^{\downarrow, \hspace{1pt}0}_{\hspace{1pt}0, \Downarrow}\right)$) states and correspond to the observed anticrossings indicated by callouts 5 and 6 in Fig.~\ref{ExptHoleMix}e. The explanation of all of these complex anticrossing patterns by the inclusion of a single term in the matrix Hamiltonians provides strong phenomenological evidence for the existence of hole spin mixing. In the next section we will address the physical origin of such a mixing term in the misalignment of QDs along the stacking axis.

The phenomenological matrix Hamiltonians we use have a limited capacity to make quantitative predictions in the case of hole spin mixing. In typical QDMs that do not show hole spin mixing, it is possible to measure each parameter independently in order to construct a quantitatively reasonably matrix Hamiltonian or to fit a single undetermined parameter. In this case, the hole spin mixing is of the same order of magnitude as spin-conserving tunneling, electron-hole exchange and the resonant change in g factor. Consequently it is impossible to determine each parameter independently or to obtain a quantitative fit to all parameters simultaneously. This limitation manifests in the suppression of the spin-conserving tunneling rate $t_{X^0}$. Lateral offset between the QDs is expected to suppress the tunneling rate and the reversal of molecular orbitals predicts that the tunneling rate should be very small for barrier thicknesses near those of this sample.\cite{DotyPRL09} A very small tunneling rate is therefore entirely plausible but the limitations of our model do not allow us to be more quantitative.

We conclude this section with a discussion of the extent of the experimental evidence for the presence of the hole spin mixing term. We have observed PL patterns evidencing hole spin mixing in the neutral exciton spectra of 6 other QDMs for samples with barrier thicknesses of 3 and 4nm. Preliminary work indicates that the effect is much smaller in a sample with a 6nm barrier. The experimental data presented in Fig.\ref{ExptHoleMix}b and e provides a striking and clear example of the contributions of the hole spin mixing term because the tunneling term is unusually small. This small tunneling rate is consistent with, but not proof of, a lateral misalignment between dots.\footnote{We observe that the coupling of the atomic ground state of the bottom dot with the first atomic excited state of the top dot leads to a larger anticrossing than the coupling of the two ground states. This is also consistent with a lateral misalignment of the dots, as the first atomic excited state has a p-like envelope function that could couple more easily to a misaligned bottom dot than the s-like envelope of the atomic ground state.}

We find additional experimental confirmation of the existence of a hole spin mixing term in the magneto-PL spectra of the positively charged trion (not shown). The excited state of the positive trion contains two hole spins, which can be in a triplet configuration when the holes are in separate dots. The hole spin mixing again introduces new anticrossings in the experimental spectra, in this case between the triplet states and singlet states that have both holes in the same dot. The additional anticrossings that appear in both the neutral exciton and positive trion spectra are enabled by the hole spin mixing term, which allows spin-flip tunneling. The spin-flip tunneling can be seen in the phenomenological Hamiltonian (Eq.~(\ref{MatrixHamiltonian})): the hole spin mixing term connects eigenstates that have holes in different dots with different spin orientations.

\section{Physical origin of the hole spin mixing term}\label{origin}
Although we can model the observed data with the addition of a phenomenological spin mixing term, the question of the physical origin of such a term remains. In analogy with single QDs, one might suspect Rashba and Dresselhaus SO interactions.\cite{BulaevPRL05,HeissPRB07}
However the large magnitude of the spin anticrossing gap in Fig.~\ref{ExptHoleMix}, $\Delta \sim 180$ $\mu$eV,
is not consistent with the high spin purity reported for holes in single QDs.\cite{HeissPRB07,GerardotNature08}
Instead, it seems that the QDM geometry has enabled a new mechanism leading to strong spin mixing. We propose that such a mechanism is the SO interaction mediated by LH. In bulk semiconductors LH are known to couple HH states with orthogonal spins.\cite{Luttinger55}
This effect is small in single QDs because LH are high in energy.\cite{BulaevPRL05} In QDMs, however, the small effective mass of LH causes them to have large tunneling rates. As a result, bonding LH states are close in energy to the lowest-lying HH states,\footnote{For sufficiently strong tunneling they may even replace HH as the ground state, see \cite{ZhuSmall09}.}
and their influence becomes important.

\begin{figure}[t]
\includegraphics[width=8.6cm]{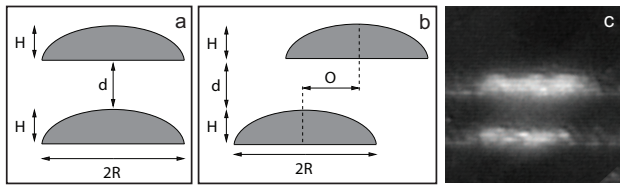}
\caption{Schematic depiction of lens-shaped QDs in an ideal QDM with no misalignment (a) and a QDM with misalignment (b). c) Cross-sectional STM image of vertically stacked InAs QDs showing misalignment.}
\label{misalign}
\end{figure}

To study the effect of the valence band SO interaction, we use the simplest description of hole states including HH-LH coupling: the four-band Luttinger-Kohn Hamiltonian.\cite{Luttinger55} The solutions of this Hamiltonian are Luttinger spinors, four component objects with two HH and two LH components. The expression of a Luttinger spinor in an ideal QDM, formed by two identical lens-shaped QDs
perfectly aligned along the stacking axis (left panel of Fig.~\ref{misalign}) is:\cite{ClimentePRB08,RegoPRB97}

\begin{equation}
|F_z,k \rangle =
\left(
\begin{array}{l}
c_{\scriptscriptstyle{+\frac{3}{2}}}\, f_{m_z}(\mathbf{r})\,    |J_z=+\frac{3}{2}\rangle \\
c_{\scriptscriptstyle{-\frac{1}{2}}}\, f_{m_z+1}(\mathbf{r})\, |J_z=+\frac{1}{2}\rangle \\
c_{\scriptscriptstyle{+\frac{1}{2}}}\, f_{m_z+2}(\mathbf{r})\,  |J_z=-\frac{1}{2}\rangle \\
c_{\scriptscriptstyle{-\frac{3}{2}}}\, f_{m_z+3}(\mathbf{r})\, |J_z=-\frac{3}{2}\rangle \\
\end{array}
\right).
\label{eq_spinor}
\end{equation}

\noindent Here $|J_z>$ is the Bloch part of the wavefunction, $f(\mathbf{r})$ is the envelope function,
and $c_{J_z}$ a numerical coefficient which gives the weight of each component.
The envelope components of the spinor have the symmetries of the confining potential.
Since the ideal QDM has circular symmetry, we can label each of the components by
an envelope angular momentum $m_z=0,\pm 1,\pm 2\ldots$.
The complete Luttinger spinor, however, does not have circular symmetry because
it is broken by the valence band SO interaction.
Instead, the spinor can be classified by the total angular momentum
$F_z=m_z+J_z$ and the main quantum number $k$.

As can be seen in Eq.~(\ref{eq_spinor}), the spinor contains a mixture of $\Uparrow$ ($J_z=+3/2$) and $\Downarrow$ ($J_z=-3/2$) HH components. Despite this mixture, the low-lying hole states of a QD ($|F_z=\pm 3/2\rangle$) are dominated by the HH with $m_z=0$. In typical InAs QDMs, one HH spin component makes up over $95\%$ of the ground state weight, with small contributions from the LH components (less than $5\%$) and the HH component with opposite spin (less than $0.1\%$). One can then identify the $|F_z=+3/2\rangle$ and $|F_z=-3/2\rangle$ spinors with the $\Uparrow$ and $\Downarrow$ HH of the usual single-band description.

\begin{figure}[h]
\includegraphics[width=8.6cm]{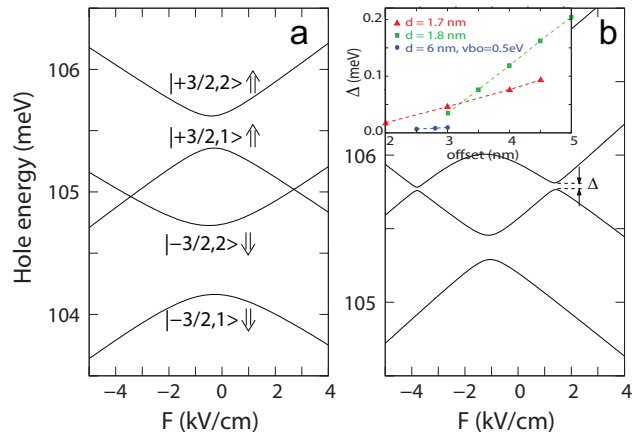}
\caption{(Color online). Hole energy levels vs.~electric field at B = 6 T, calculated with k$\cdot$p theory:
(a) no misalignment; (b) misaligned QDM with 3 nm lateral offset.
In (a) we indicate the $|F_z,k\rangle$ quantum numbers and the pseudo-spin.
Note that the states with opposite spin cross (anticross) in the absence (presence) of misalignment.
The inset in (b) shows the magnitude of the spin anticrossing gap as a function of the lateral offset
for various interdot barriers.}
\label{misalign2}
\end{figure}

To show that the weak mixture of HH spins within the spinor of a circularly symmetric QDM is not responsible for the features observed in Fig.~\ref{ExptHoleMix}, we calculate the low-energy hole states of an ideal InAs/GaAs QDM subject to a longitudinal magnetic field of $B=6$ T (see Appendix \ref{app:theo} for calculation details). The resulting energy spectrum is plotted in Fig.~\ref{misalign2}a. The $|F_z,k\rangle$ symmetry of the states is indicated, along with the spin of the dominant HH component of each spinor. In the figure we see anticrossings at resonant electric fields ($E_z \sim 0$ kV/cm), where states with the same $F_z$ mix to form bonding and antibonding molecular states. These anticrossings correspond to the spin-conserving tunneling observed in typical experimental spectra. In addition, Fig.~\ref{misalign2}a shows crossings between levels with different $F_z$ (different pseudo-spin). These states cross (rather than anticross) because the SO interaction does not mix states with different $F_z$. These crossings, however, occur between the states that anticross in both the experimental spectra and the phenomenological calculations that include the spin mixing term $hm$ (Fig.~\ref{ExptHoleMix}). The absence of anticrossings at these points in Fig.~\ref{misalign2}a demonstrates that the inclusion of the SO interaction is not sufficient to explain the new experimentally observed anticrossings.

In order to obtain anticrossings that match the experimental spectra, the $F_z$ symmetry must be broken. The total angular momentum symmetry can be removed by structural distortions breaking the circular symmetry of the QDM. The distortion could be dot eccentricity, which is often present in Stranski-Krastanov grown QDs.\cite{GarciaAPL97} However, our simulations (not shown) indicate that eccentricity only weakly mixes states with different $F_z$, and the anticrossings gaps it produces never reach the large experimental value. Instead we consider a lateral offset between the QDs which form the QDM, as shown in Fig.~\ref{misalign}b. QDM misalignment efficiently removes the circular symmetry at resonant electric fields, leading to a strong mixing of states with different $F_z$ and, as we show below, to spin anticrossing gaps comparable to those of the experiment.

The effect of misalignment is illustrated in Fig.~\ref{misalign2}b, where we plot the hole
energy spectrum for the same QDM as in Fig.~\ref{misalign2}a, but now including a lateral offset of 3 nm.
One can see that the presence of misalignment introduces the expected anticrossings between
states with opposite pseudo-spin.

The inset in Fig.~\ref{misalign2}b shows the magnitude of the anticrossing gap between states
with different pseudo-spin, $\Delta$, as a function of the lateral offset.
InAs QDMs with different interdot barrier thickness are considered.\footnote{The misalignment
modifies not only the magnitude of the spin-flip tunneling rate, but also that of the spin conserving one.
Therefore, for a given barrier thickness and lateral offset, the states with opposite pseudo-spin may not
intersect. This is why in Fig.~\ref{misalign2}b inset we only plot segments.}
$\Delta$ increases linearly with the offset, and for 5 nm it may reach values
of 200 $\mu$eV, which are comparable to the value observed in the magneto-photoluminescence
spectra of Fig.~\ref{ExptHoleMix}.
The inset also reveals that the effect of the barrier thickness (d) and height (valence band offset) is important.
In general, the weaker the tunneling the smaller the anticrossing gap (compare e.g. d=$1.7$ nm
and d=6 nm). This is because the bonding LH states are farther in energy, and their influence decreases.
The nature of the hole molecular state is also relevant. At d=$1.7$ nm the calculated
ground state is bonding, but it switches to antibonding at d=$1.8$ nm. The antibonding ground state contains a larger admixture of spinor components\cite{ClimentePRB08}, which explains the drastic increase in anticrossing gap as a result of a small increase in barrier thickness.

\begin{figure}[h]
\includegraphics[width=8.6cm]{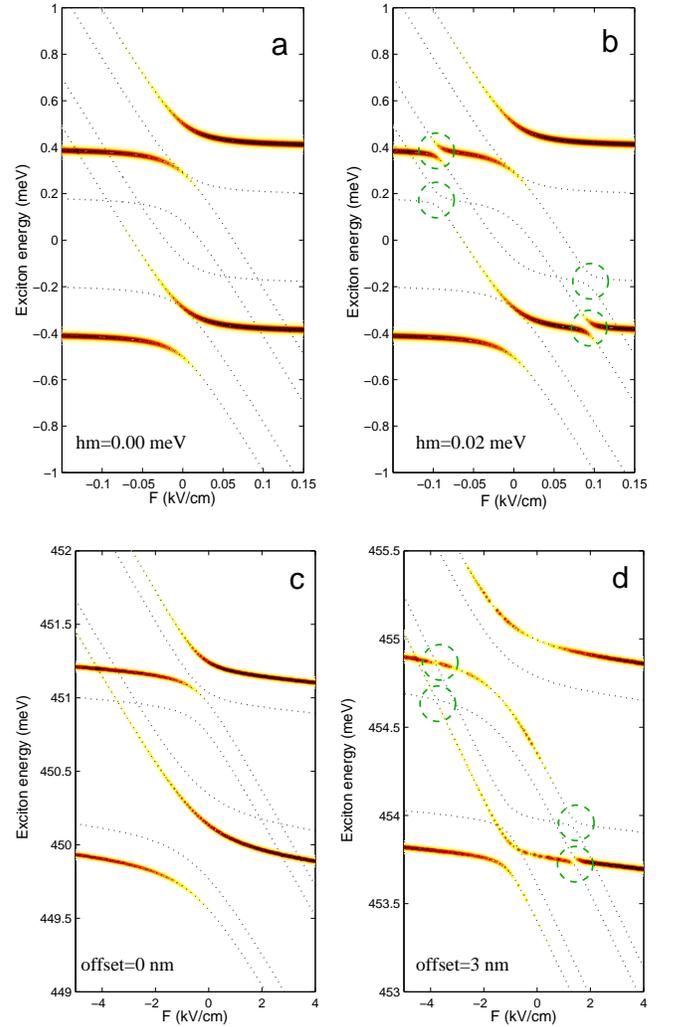}
\caption{(Color online). Exciton energy levels (dotted lines) and photoluminescence spectrum at B = 6 T
calculated with Eq.~(\ref{MatrixHamiltonian}) (a-b), and k$\cdot$p theory (c-d).
Note the correspondence between (a) and (c), and between (b) and (d).
The presence of QDM misalignment originates the same anticrossings as the spin mixing term $hm$
(highlighted by dashed circles).}\label{misalign3}
\end{figure}

To further support misalignment as the origin of the phenomenological spin mixing term of Eq.~(\ref{MatrixHamiltonian}) we show that the inclusion of lateral offset in the k$\cdot$p theory introduces anticrossings between exactly the same states that anticross when the spin mixing term is included in the phenomenological Hamiltonian. In Fig.~\ref{misalign3} we compare the exciton emission spectrum calculated with the phenomenological Hamiltonian (top row) and k$\cdot$p theory (bottom row). The parameters of the phenomenological Hamiltonian are the same as in Fig.~\ref{ExptHoleMix}, but the resonant $g$-factor and electron-hole exchange have been neglected for simplicity. Because these terms have been neglected, the results should not be compared with Fig.~\ref{ExptHoleMix}f. Fig.~\ref{misalign3}(a) and (c) correspond to the system with $hm=0$ (phenomenological) and no lateral offset (k$\cdot$p). Panels (b) and (d) correspond to the system with $hm=0.02$ meV and 3 nm offset, respectively. As highlighted by the dashed circles, the inclusion of finite $hm$ and finite offset introduces anticrossings at the same positions. This strongly supports our conclusion that lateral offset is responsible for the phenomenological spin mixing term.

\begin{figure*}[t]
\centering
\includegraphics[width=16.8cm]{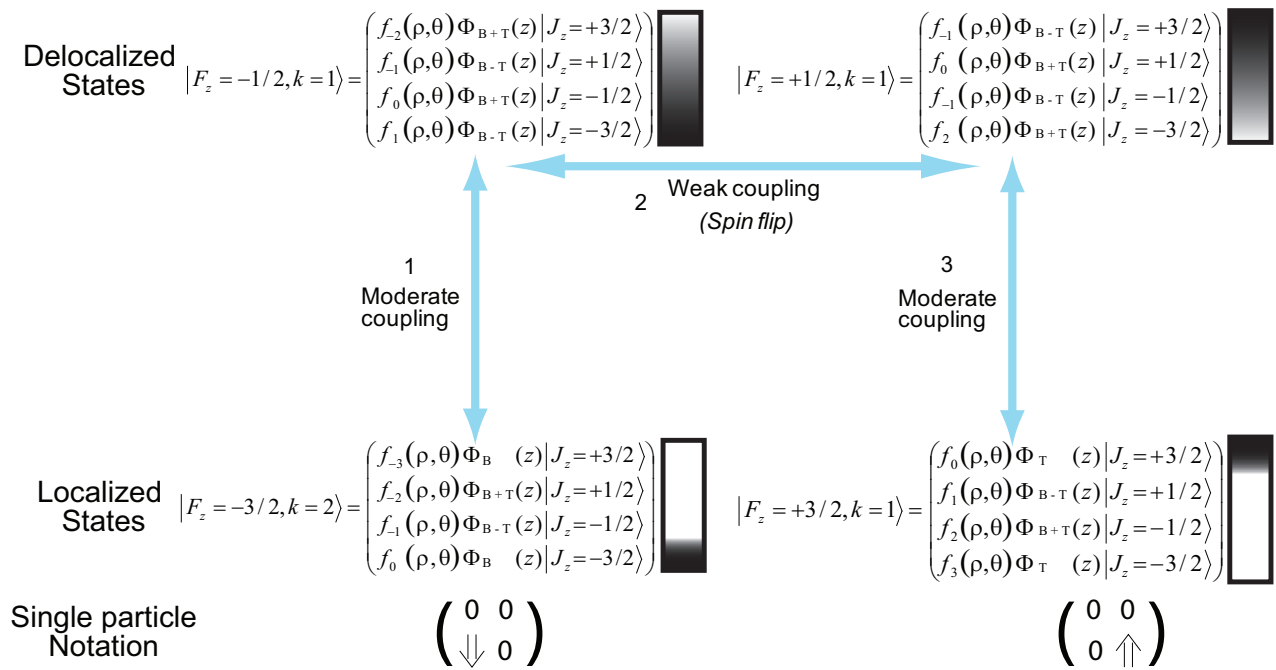}
\caption{(Color online) Diagrammatic representation of the coupling between two localized ``HH'' states with opposite pseudo-spin and location, as indicated by the single particle notation. All four spinor components of each $|F_z,k\rangle$ quantum state are shown, with the shaded boxes indicating the approximate relative weights of each component. The interaction between localized `HH' states is mediated by delocalized states with large LH components. Note that the spinor labels are only approximate: misalignment breaks the symmetry, mixing different spinor configurations and enabling hole spin mixing.}\label{CouplingSchematic}
\end{figure*}

A detailed derivation of the spin mixing term $hm$ in terms of the k$\cdot$p matrix elements induced by
the misaligment potential can be found in Appendix \ref{app:theo}. It follows from the analysis that
the proximity of LH states plays a critical role in the spin mixing of HH.
The direct coupling between $|3/2,1\rangle$ and $|-3/2,2\rangle$ states is small,
because these two states are essentially HHs localized in opposite dots at the electric fields
where the intersections occur. The coupling is mostly mediated by the excited states with $F_z=\pm 1/2$, which contain sizable LH components and are hence delocalized over the QDM for the entire range of electric
fields under study. This allows them to couple the two ``HH'' states efficiently. A diagram summarizing the coupling is shown in Fig.~\ref{CouplingSchematic}. To illustrate the localization of the states, in-plane and
vertical parts of the spinor envelope components are written separately. $B$ ($T$) indicates localization in the bottom (top) dot, while ($B \pm T$) indicates bonding and antibonding delocalized states. Mixing occurs between states with the same $J_z$. The coupling between the ``HH'' and delocalized ``LH'' states is moderately strong. The hole spin mixing is limited by the coupling between the two delocalized ``LH'' states because these two spinors accumulate weight on opposite components.

To close this section we mention that the existence of a lateral offset between the QDs forming the QDM is quite plausible. In the right panel of Fig.~\ref{misalign} we show cross-sectional scanning tunneling microscopy (XSTM) data of vertically stacked InAs QDs showing a lateral offset along the stacking axis. It is difficult to get good statistics on the occurrence of lateral offsets from XSTM data because it is sensitive only to lateral offsets parallel to the cleavage plane. A survey of our XSTM data indicates that lateral offsets are not uncommon in our QDMs. However, the lateral offsets are typically much smaller than the 5 nm offset found to fit the spectra presented in Fig.~\ref{ExptHoleMix}. This is consistent with the observation that the degree of hole spin mixing in this example is unusually large.

\section{Summary}
\label{conclusion}
We observe optical intensity from dark excitons in a number of QDM samples. This dark exciton optical intensity can be phenomenologically explained by the presence of hole spin mixing. We have presented an example in which the hole spin mixing is sufficiently large and the tunneling sufficiently small that bright-dark exciton anticrossings can be directly observed and measured. We have used a phenomenological matrix Hamiltonian to demonstrate that the inclusion of hole spin mixing qualitatively explains all of the new anticrossings in the experimental data. We have then used a k$\cdot$p theory to show that molecular symmetry breaking in the form of QD lateral offset changes the mixing of HH and LH states and enables the same form of hole spin mixing. The combination of experimental evidence and k$\cdot$p theory lead us to the conclusion that hole spin mixing can occur in QDMs as a result of symmetry breaking.

A quantitative understanding of the spin mixing term will require detailed experiments in which the molecular symmetry can be broken in a quantifiable way. It will also require the development of more sophisticated theory and modeling techniques to quantitatively determine the magnitude of spin mixing and its dependence on molecular structure and symmetry. The experimental evidence and k$\cdot$p theory we present suggest that there is a rich regime of spin physics to be explored in QDMs. In the final section we discuss the implications of hole spin mixing for device applications.

\section{Implications of hole spin mixing}
\label{Sect:Implications}
\begin{figure}[h]
\includegraphics[width=6.5cm]{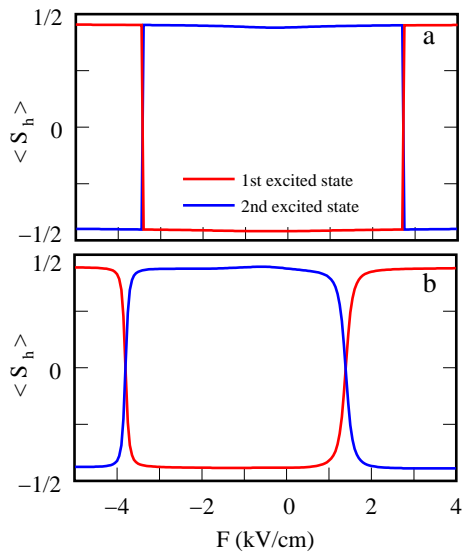}
\caption{(Color Online) Expectation value of the hole pseudo-spin as a function of applied electric field. Red and blue lines correspond to the first and second excited states of Fig.~\ref{misalign2}. (a) no lateral offset. (b) 3 nm lateral offset. The reversal of red and blue lines in panel (a) arises when the energy order of the two states reverses and does not indicate a degradation of hole spin purity.}
\label{HoleSpinPurity}
\end{figure}

As discussed earlier, the low-lying hole states with $F_z=\pm 3/2$ have a strongly dominant HH component and hence can be assigned a pseudo-spin. However, at the electric fields where the anticrossings induced by misalignment occur, these states mix with other spinors with opposite pseudo-spin. To quantify the effect of such mixing on the spin purity, in Fig.~\ref{HoleSpinPurity} we plot the expectation value of the hole pseudo-spin $S_h$ for the first and second molecular excited states of Fig.~\ref{misalign2} (i.e. the states that are a mixture of $|F_z=+3/2,1\rangle$ and $|F_z=-3/2,2\rangle$) as a function of the applied electric field.
$\langle S_h \rangle$ is evaluated from the weight of the components of the Luttinger spinor as

\begin{equation}
\langle S_h \rangle = \frac{1}{2}\,\left( c_{\scriptscriptstyle{+\frac{3}{2}}}^2 + c_{\scriptscriptstyle{+\frac{1}{2}}}^2/3
- c_{\scriptscriptstyle{-\frac{1}{2}}}^2/3 -  c_{\scriptscriptstyle{-\frac{3}{2}}}^2 \right),
\end{equation}

\noindent where the factor $1/3$ acting upon the LH components comes from the Bloch function coefficients.
Fig.~\ref{HoleSpinPurity}a and Fig.~\ref{HoleSpinPurity}b correspond to the QDM without and with lateral offset,
respectively (as in Figs.~\ref{misalign2}a and \ref{misalign2}b).
One can see that in the absence of misalignment $\langle S_h \rangle$ is nearly pure ($\pm 1/2$),\footnote{$\langle S_h\rangle$ would
be even closer to $\pm 1/2$ if we included strain in our model, further reducing the LH contribution to the low-lying spinors.}
but the inclusion of misalignment severely degrades the spin purity at electric fields near the spin anticrossing points.

These perturbations in the hole spin purity further illustrate that the HH-only approximation has only limited validity in QDM. Near the anticrossing points, the effect of valence band mixing must be taken into account when designing information storage or manipulations protocols based on the hole spin projections. The contributions of the other spinor components may enable additional decoherence or dephasing mechanisms that limit the ability to store and manipulate quantum information, as well as the preparation of pure spin hole states through the positive trion\cite{GerardotNature08,EbbensPRB05}. These mechanisms could be suppressed by minimizing the structural distortions that break the molecular symmetries or by designing spin storage and manipulation protocols that are insensitive to the mixing of additional hole spin components. Away from the anticrossing points, the HH spin projections remain relatively pure and viable for quantum information storage.

The spin mixing mechanism we describe here is reminiscent of that reported by Ferreira and Bastard for single
asymmetric quantum wells.\cite{FerreiraEPL93} Holes in symmetric quantum wells have a well defined pseudo-spin in spite of the HH-LH coupling because the parity symmetry prevents undesired mixing between spinors.\cite{AndreaniPRB87, RegoPRB97} Introducing envelope function asymmetries that break the parity bring about a D'yakanov-Perel-like mechanism of spin relaxation.\cite{ZuticRMP04} In our QDM system, the vertical parity symmetry is lifted by the lens-shaped confinement and the different composition of the dots. Still, this does not suffice to mix pseudo-spins because the lateral confinement, which is obviously absent in quantum wells, imposes an additional symmetry, namely circular symmetry. As a result, the ``spin up'' and ``spin down'' hole states have different total angular momenta $F_z$ and they remain orthogonal. In order to mix the orthogonal spin projections, one has to break the rotational symmetry. As we have shown, this is most efficiently achieved by the misalignment of the QDM.

The envelope origin of the SO term we report suggests that it can be controlled with external
field, and hence used in a similar fashion to the Rashba term for spintronic applications.\cite{ZuticRMP04}
Indeed, the magnitude of the spin anticrossing gaps observed in Fig.~\ref{ExptHoleMix}(e) are comparable
to those of electrons in InAs QDMs with strong Rashba interaction.\cite{PfundPRB07} This makes holes in QDMs particularly suitable for electric-field-induced spin manipulation. For example, two orthogonal spin states of the hole in the bottom QD could provide the qubit basis. Away from the electric field of tunnel coupling and hole spin mixing, these states are well isolated and could provide a robust means of storing quantum information.\cite{HeissPRB07, GerardotNature08} The applied electric field could then be varied to bring the dots into the regime where hole spin mixing becomes strong in order to mix the two spin configurations. A detailed examination of the interaction strengths and electric field pulse sequences required to effect a single qubit rotation are beyond the scope of this paper. If feasible, however, such a scheme would require fewer resources to implement spin control protocols than current proposals. Spin rotations could be implemented with only a single applied electric field, eliminating the need for pulsed lasers tuned to transitions specific to each dot or GHz frequency fields used to implement g-TMR rotations of spin projections.\cite{EconomouPRL07, EconomouPRB08, KatoScience03, AndlauerPRB09} At the same time, the QDM structure preserves the opportunity to use optics for spin initialization, readout and control of 2-qubit operations.

\begin{acknowledgments}
We wish to thank J. Planelles and M.  Scheibner for helpful discussions.
Support from NSA/ARO, ONR, MCINN project CTQ2008-03344
and the Ramon y Cajal program (JIC) is acknowledged. MFD began work on this project while a NRC Research Associate at the Naval Research Laboratory.
\end{acknowledgments}

\appendix

\section{Empirical parameters}
\label{app:params}

The numerical values of the parameters used in the matrix Hamiltonian are not free fitting parameters. The value of each parameter can be determined from experimental data. $t_{X^0} = -0.107$ meV is determined from the measured zero magnetic field anticrossing. $g_{hB} = -1.695$ is determined
from the Zeeman splitting of the lower energy line well away from
the anticrossing region. Similarly, $g_{hT} = -1.66$ is determined from the
Zeeman splitting of the higher energy indirect line well away from
anticrossing. $g_{e} = -0.6$ and $\delta_0 = 0.101$ meV are determined from the
asymptotic energies of the dark states. $d = 5.8$ nm is
determined from the slope of the indirect transition energy. B = 6 T is
the known value of the applied magnetic field. $\mu_B = 0.0579$ meV is the Bohr magneton. $g_{12}=0.47$ is fit to
the data by looking at the higher energy anticrossing, which shows
little affect from the hole mixing. $hm= 0.092$ meV in the calculations of Fig.\ref{ExptHoleMix}c and f is fit by looking at the new bright-dark anticrossings. The suppression of $t_{X^0}$ in the calculations with nonzero $hm$ is intuitively expected because the lateral offset increases the distance between the center of the QDs. k$\cdot$p calculations confirm that the LH influence does not alter the suppression of $t_{X^0}$. Note again that the calculations in Fig.\ref{ExptHoleMix} use both electron spin projections, so the matrix and intensity vector analogous to Eqn.(\ref{MatrixHamiltonian}) for the electron spin up case is also used. A value of $Ind = 0.5$ is used to generate the calculated spectral maps.

\section{Theoretical model and k$\cdot$p derivation of spin mixing terms}
\label{app:theo}

In this Appendix we describe the theoretical model employed for the
k$\cdot$p calculations and then derive an expression
for the spin mixing operator of Eq.~(\ref{MatrixHamiltonian}), $hm$,
in terms of the matrix elements induced by misalignment.

We write the Hamiltonian of the QDM as:

\begin{equation}
\label{eq1app}
{\cal H}={\cal H}_{\mbox{\scriptsize LK}} + V_{\mbox{\scriptsize QDM}} + V_{\mbox{\scriptsize offset}}.
\end{equation}

\begin{figure}[h]
\includegraphics[width=8.6cm]{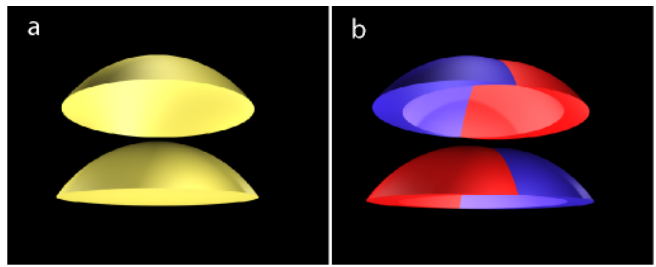}
\caption{(Color Online) (a) Confinement potential of an ideal QDM.
(b) Perturbative potential introduced by QDM misalignment.
Blue and red regions correspond to $+V_o$ and $-V_o$, respectively.}
\label{pots}
\end{figure}

\noindent Here ${\cal H}_{\mbox{\scriptsize LK}}$ is the three-dimensional
Luttinger-Kohn Hamiltonian, including longitudinal magnetic and electric fields.
$V_{\mbox{\scriptsize QDM}}$ is the
confinement potential of an ideal QDM, formed by two vertically stacked lens-shaped
QDs (spherical caskets) with perfect alignment (see Fig.~\ref{pots}a).
The potential is zero inside the dots and $V_o$ outside.
$V_{\mbox{\scriptsize offset}}$ is a perturbative
potential induced by laterally offsetting the QDs. It is the difference between
the potential of misaligned and aligned QDM (see Fig.~\ref{pots}b).

We first obtain the eigenstates of
${\cal H}_{\mbox{\scriptsize sym}}={\cal H}_{\mbox{\scriptsize LK}} + V_{\mbox{\scriptsize QDM}}$
and then project the full Hamiltonian ${\cal H}$ into these states.
The minimal basis set which captures the spin mixing features of Eq.~(\ref{MatrixHamiltonian}) is formed by
the six lowest-lying eigenstates of ${\cal H}_{\mbox{\scriptsize sym}}$, namely:

\begin{align*}
| 1 \rangle = | F_z = -\frac{3}{2}, k=1 \rangle,& & | 2 \rangle &= | F_z = -\frac{3}{2}, k=2 \rangle, \\
| 3 \rangle = | F_z = -\frac{1}{2}, k=1 \rangle,& & | 4 \rangle &= | F_z = +\frac{1}{2}, k=1 \rangle, \\
| 5 \rangle = | F_z = +\frac{3}{2}, k=1 \rangle,& & | 6 \rangle &= | F_z = +\frac{3}{2}, k=2 \rangle.
\end{align*}

The above states are obtained by numerically integrating the Luttinger-Kohn Hamiltonian,
written in cylindrical coordinates, with a finite difference scheme.\cite{PlanellesPRB01}
The complete Hamiltonian ${\cal H}$ is then integrated using an exact diagonalization technique.
The QDM is constructed of two QDs with radius $15$ nm, height $2$ nm
and an interdot separation of $d=1.7$ nm (unless otherwise stated).
We use InAs Luttinger parameters $\gamma_1=20.0$, $\gamma_2=8.5$ and $\gamma_3=9.2$.\cite{VurgaftmanJAP01}
The valence band offset is $V_o=200$ meV, and a constant g-factor $g_h=-1.5$ is assumed.

The energy of the $F_z=\pm 3/2$ states as a function of the electric field
is represented in Fig.~\ref{misalign2}a for an ideal QDM at B = 6 T.
The $F_z=\pm 1/2$ states are a few meV higher in energy. % ~ 10 meV
From the dominant component of the spinor, one can identify the states
$|1\rangle$ and $|2\rangle$ ($|5\rangle$ and $|6\rangle$)
with the $\Downarrow$ ($\Uparrow$) HHs of Eq.~(\ref{atomic_basis}).
These states are localized either in the top or the bottom dot,
except for a narrow window near the resonant field $E_z \sim 0$ where they
form delocalized bonding and antibonding states.
On the other hand, the excited states $|3\rangle$ and $|4\rangle$ have a significant
admixture of HH and LH components. This allows them to be significantly delocalized
for all the values of $E_z$ in the figure, forming bonding molecular states.\footnote{Due
to the strong tunneling of LHs, the corresponding antibonding states are much higher in energy
and have neglegible influence on the spin mixing.} %although they may modify the spin-conserving tunneling rate.

The exciton calculations of section \ref{origin} are carried out using the hole states
calculated as described above and electron states calculated in a similar fashion but with
a single-band effective mass model.\cite{PlanellesPRB01} The electron mass is $m^*=0.06$,
the conduction band offset is set to $V_o=500$ meV and the g-factor $g_e=-0.6$.
To mimic the experimental situation, we force the electron to stay in the bottom QD
(a single dot potential is used). Electron-hole Coulomb interaction is accounted for using
a configuration interaction method on the basis of the Hartree products formed by the electron
ground state and the six lowest hole states. The matrix elements are integrated using Monte-Carlo
routines, and the exciton emission intensity is computed within the dipole
approximation.\cite{Jacak98} \\

We can obtain a perturbative expression for the spin mixing term $hm$ by projecting
Hamiltonian (\ref{eq1app}) into the basis of the states $|1\rangle$ to $|6\rangle$.
This yields:

\begin{equation}
\label{eq4app}
\begin{array}{cccccccc}
\left(\begin{array}{cccccc}
E_1 & V_{12} & V_{13}     & 0  &  0  &  0  \\
V_{12}^* & E_2 & V_{23}   & 0  &  0  &  0  \\
V_{13}^* & V_{23}^* & E_3 & V_{34}  &  0  &  0  \\
0  &  0  &  V_{34}^* & E_4 & V_{45} &  V_{46}  \\
0  &  0  &  0        & V_{45}^* & E_5 & V_{56} \\
0  &  0  &  0        & V_{46}^*  &  V_{56}^* & E_6 \\
\end{array}\right)
\end{array}
\end{equation}

\noindent where $E_i$ is the energy of the hole state $|i\rangle$, while $V_{ij}=\langle i| V_{\mbox{\scriptsize offset}} |j \rangle$
is the matrix element induced by the misalignment potential, which couples the states $|i\rangle$ and $|j\rangle$.

For simplicity, in Eq.~(\ref{eq4app}) we have set to zero the matrix elements which are small or non relevant
to the spin mixing under discussion.
In particular, we note that there is no direct mixing between the HH $\Downarrow$ and $\Uparrow$ states.
This can be understood from an analysis of the matrix elements. For example,
$\langle 1 | V_{\mbox{\scriptsize offset}} | 5 \rangle$ reads:

\begin{equation}
%\langle 1 | V_{\mbox{\scriptsize offset}}| 5 \rangle \,=\,
\left \langle
\begin{array}{l}
c_{\scriptscriptstyle{+\frac{3}{2}}} '\, f_{-3}(\mathbf{r}) \langle J_z=+\frac{3}{2}| \\
c_{\scriptscriptstyle{-\frac{1}{2}}} '\, f_{-2}(\mathbf{r}) \langle J_z=+\frac{1}{2}|\\
c_{\scriptscriptstyle{+\frac{1}{2}}} '\, f_{-1}(\mathbf{r}) \langle J_z=-\frac{1}{2}|\\
c_{\scriptscriptstyle{-\frac{3}{2}}} '\, f_{+0}(\mathbf{r})  \langle J_z=-\frac{3}{2}|\\
\end{array}
\right |
V_{\mbox{\scriptsize offset}} \, {\cal I}
\left |
\begin{array}{l}
c_{\scriptscriptstyle{+\frac{3}{2}}}\, f_{0}(\mathbf{r}) | J_z=+\frac{3}{2} \rangle \\
c_{\scriptscriptstyle{-\frac{1}{2}}}\, f_{1}(\mathbf{r}) | J_z=+\frac{1}{2} \rangle \\
c_{\scriptscriptstyle{+\frac{1}{2}}}\, f_{2}(\mathbf{r}) | J_z=-\frac{1}{2} \rangle \\
c_{\scriptscriptstyle{-\frac{3}{2}}}\, f_{3}(\mathbf{r}) | J_z=-\frac{3}{2} \rangle \\
\end{array}
\right \rangle,
\nonumber
\end{equation}

\noindent where ${\cal I}$  is the identity matrix.
While $|1\rangle$ gathers most of its weight in the $J_z=-3/2$ component, which has the lowest
envelope angular momentum ($m_z=0$), $|5\rangle$ does so in the $J_z=+3/2$ component.
Since each component of the bra couples to that of the ket with equal $J_z$,
the coupling between the two vectors is negligible.
By contrast, both $|1\rangle$ and $|5\rangle$ couple with the excited states
$|3\rangle$ and $|4\rangle$. This is because such states have a strong admixture of
components and are delocalized over the entire QDM, so they are able to couple with
``HH'' states of either ``spin'' regardless of their localization.

In order to compare with Eq.~(\ref{MatrixHamiltonian}), we reduce Hamiltonian (\ref{eq4app}) to
an effective 4$\times$4 Hamiltonian on the basis of ``HH'' states only (i.e. states $|1\rangle$, $|2\rangle$, $|5\rangle$ and $|6\rangle$). After some algebra, we obtain:

\begin{widetext}
%{\small
\begin{equation}
\label{eq7app}
\begin{array}{cccccc}
\left(\begin{array}{cccc}
E_1 + k  (\lambda - E_4)   V_{13} V_{13}^* & V_{12} + k  (\lambda-E_4)  V_{13}  V_{23}^* & k V_{34}^*  V_{13}  V_{45}^* & k  V_{34}^*  V_{13}  V_{46}^*\\
V_{12}^* + k (\lambda-E_4) V_{13}^* V_{23} & E_2 + k  (\lambda - E_4)   V_{23} V_{23}^*  & k V_{34}^*  V_{23}  V_{45}^* & k  V_{34}^*  V_{23}  V_{46}^*\\
k V_{34} V_{13}^* V_{45}  &  k V_{34} V_{23}^* V_{45} &  E_5 + k (\lambda - E_3) V_{45} V_{45}^*   & V_{56} + k (\lambda-E_3) V_{45} V_{46}^* \\
k V_{34} V_{13}^* V_{46}  &  k V_{34} V_{23}^* V_{46} & V_{56}^* + k (\lambda-E_3) V_{45}^* V_{46} & E_6 + k (\lambda - E_3) V_{46} V_{46}^*  \\
\end{array}\right)
\end{array}
\end{equation}
%}
\end{widetext}

\noindent where $\lambda$ are Hamiltonian (\ref{eq4app}) eigenvalues and
$k= \left( (\lambda-E_3)(\lambda-E_4) - V_{34} V_{34}^* \right)^{-1}$.

Hamiltonian (\ref{eq7app}) can be simplified if we disregard the imaginary part of the matrix elements,
which is neglegible in our QDM because $V_{\mbox{\scriptsize offset}}$ has even parity along
the azimuthal direction, and we further consider that the orbital part of
$|1\rangle$ and $|5\rangle$ ($|2\rangle$ and $|6\rangle$) are very similar, as they differ in the pseudo-spin only.
It follows that $V_{12} \approx V_{56}=a$, $V_{13} \approx V_{45} = \alpha$ and $V_{23} \approx V_{46} = \beta$.
The Hamiltonian can then be rewritten as:

\begin{widetext}
\begin{equation}
\label{eq9app}
\begin{array}{cccccc}
\left(\begin{array}{cccc}
%E_1 + k  (\lambda - E_4)   V_{13}^2 & V_{12} + k  (\lambda-E_4)  V_{13}  V_{23} & hm_1 & hm  \\
%V_{12} + k (\lambda-E_4) V_{13} V_{23} & E_2 + k  (\lambda - E_4)   V_{23}^2    & hm  & hm_2 \\
%hm_1  &  hm  &  E_5 + k (\lambda - E_3) V_{45}^2 & V_{56} + k (\lambda-E_3) V_{45} V_{46} \\
%hm   &  hm_2 &  k (\lambda-E_3) V_{45} V_{46} & E_6 + k (\lambda - E_3) V_{46}^2  \\
E_1 + k (\lambda - E_4) \alpha^2 & a + k  (\lambda-E_4) \alpha \beta & hm_1 & hm  \\
a + k (\lambda-E_4) \alpha \beta & E_2 + k  (\lambda - E_4) \beta^2  & hm  & hm_2 \\
hm_1  &  hm  &  E_5 + k (\lambda - E_3) \alpha^2 & a + k (\lambda-E_3) \alpha \beta \\
hm   &  hm_2 &  k (\lambda-E_3) \alpha \beta & E_6 + k (\lambda - E_3) \beta^2  \\
\end{array}\right)
\end{array}
\end{equation}
\end{widetext}

\noindent where $hm = k V_{34} \alpha \beta$, $hm_1=k V_{34} \alpha^2$ and $hm_2=k V_{34} \beta^2$.\\

\begin{figure}[h]
\begin{center}
\includegraphics[width=7.0cm]{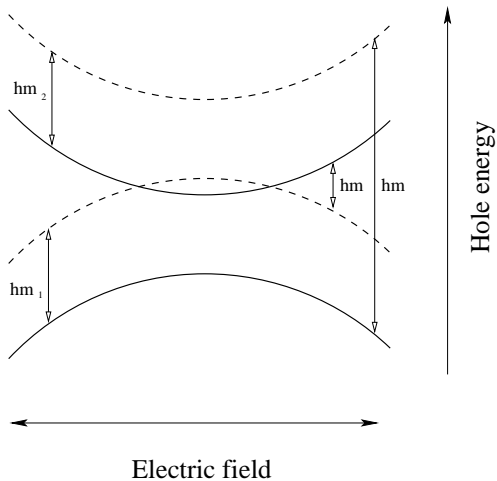}
\caption{Schematic representation of the hole energy levels of Fig.~\ref{misalign2}
indicating the level mixing induced by the spin-orbit terms $hm$, $hm_1$ and $hm_2$
of Hamiltonian (\ref{eq9app}).
Solid (dashed) lines are used for states with pseudo-spin $\Downarrow$ ($\Uparrow$).}
\end{center}\label{scheme_hm}
\end{figure}

By inspecting the effective Hamiltonian (\ref{eq9app}), a number of conclusions can be drawn.
First and foremost, we have found an expression for the spin-mixing term $hm$ in terms of
misalignment matrix elements. This clearly reveals the importance of the indirect
coupling through the excited $|F_z=\pm 1/2\rangle$ states ($V_{34}$).
Indeed, $hm$ is inversely proportional to the energy splitting between the $|F_z=\pm 3/2\rangle$
and the excited $|F_z=\pm 1/2\rangle$ states (through $k$).
Second, in addition to $hm$ there are other spin mixing terms, namely $hm_1$ and $hm_2$.
These terms couple ``HH'' states localized in the same dot.
Fig.~\ref{scheme_hm} shows an schematic representation of the low-energy hole states
of a QDM at finite magnetic field and the spin mixing terms coupling each pair of states.
$hm_1$ and $hm_2$ are not needed in Eq.~(\ref{MatrixHamiltonian})
because the states they mix are split by the Zeeman energy, which in the B = 6 T experiment
is far larger than the SO perturbation. Finally, we note that misalignment not only induces spin mixing, but also affects the ``HH''
energy levels (through the diagonal terms of (\ref{eq9app})) and the spin-conserving tunneling rates.
This can be observed in the k$\cdot$p results in Fig.~\ref{misalign2}.

\begin{thebibliography}{35}
\expandafter\ifx\csname natexlab\endcsname\relax\def\natexlab#1{#1}\fi
\expandafter\ifx\csname bibnamefont\endcsname\relax
  \def\bibnamefont#1{#1}\fi
\expandafter\ifx\csname bibfnamefont\endcsname\relax
  \def\bibfnamefont#1{#1}\fi
\expandafter\ifx\csname citenamefont\endcsname\relax
  \def\citenamefont#1{#1}\fi
\expandafter\ifx\csname url\endcsname\relax
  \def\url#1{\texttt{#1}}\fi
\expandafter\ifx\csname urlprefix\endcsname\relax\def\urlprefix{URL }\fi
\providecommand{\bibinfo}[2]{#2}
\providecommand{\eprint}[2][]{\url{#2}}

\bibitem[{\citenamefont{Bayer et~al.}(2000)\citenamefont{Bayer, Stern, Kuther,
  and Forchel}}]{BayerPRB00}
\bibinfo{author}{\bibfnamefont{M.}~\bibnamefont{Bayer}},
  \bibinfo{author}{\bibfnamefont{O.}~\bibnamefont{Stern}},
  \bibinfo{author}{\bibfnamefont{A.}~\bibnamefont{Kuther}}, \bibnamefont{and}
  \bibinfo{author}{\bibfnamefont{A.}~\bibnamefont{Forchel}},
  \bibinfo{journal}{Physical Review B} \textbf{\bibinfo{volume}{61}},
  \bibinfo{pages}{7273} (\bibinfo{year}{2000}).

\bibitem[{\citenamefont{Heiss et~al.}(2007)\citenamefont{Heiss, Schaeck, Huebl,
  Bichler, Abstreiter, Finley, Bulaev, and Loss}}]{HeissPRB07}
\bibinfo{author}{\bibfnamefont{D.}~\bibnamefont{Heiss}},
  \bibinfo{author}{\bibfnamefont{S.}~\bibnamefont{Schaeck}},
  \bibinfo{author}{\bibfnamefont{H.}~\bibnamefont{Huebl}},
  \bibinfo{author}{\bibfnamefont{M.}~\bibnamefont{Bichler}},
  \bibinfo{author}{\bibfnamefont{G.}~\bibnamefont{Abstreiter}},
  \bibinfo{author}{\bibfnamefont{J.~J.} \bibnamefont{Finley}},
  \bibinfo{author}{\bibfnamefont{D.~V.} \bibnamefont{Bulaev}},
  \bibnamefont{and} \bibinfo{author}{\bibfnamefont{D.}~\bibnamefont{Loss}},
  \bibinfo{journal}{Physical Review B} \textbf{\bibinfo{volume}{76}},
  \bibinfo{pages}{241306} (\bibinfo{year}{2007}).

\bibitem[{\citenamefont{Gerardot et~al.}(2008)\citenamefont{Gerardot, Brunner,
  Dalgarno, Ohberg, Seidl, Kroner, Karrai, Stoltz, Petroff, and
  Warburton}}]{GerardotNature08}
\bibinfo{author}{\bibfnamefont{B.~D.} \bibnamefont{Gerardot}},
  \bibinfo{author}{\bibfnamefont{D.}~\bibnamefont{Brunner}},
  \bibinfo{author}{\bibfnamefont{P.~A.} \bibnamefont{Dalgarno}},
  \bibinfo{author}{\bibfnamefont{P.}~\bibnamefont{Ohberg}},
  \bibinfo{author}{\bibfnamefont{S.}~\bibnamefont{Seidl}},
  \bibinfo{author}{\bibfnamefont{M.}~\bibnamefont{Kroner}},
  \bibinfo{author}{\bibfnamefont{K.}~\bibnamefont{Karrai}},
  \bibinfo{author}{\bibfnamefont{N.~G.} \bibnamefont{Stoltz}},
  \bibinfo{author}{\bibfnamefont{P.~M.} \bibnamefont{Petroff}},
  \bibnamefont{and} \bibinfo{author}{\bibfnamefont{R.~J.}
  \bibnamefont{Warburton}}, \bibinfo{journal}{Nature}
  \textbf{\bibinfo{volume}{451}}, \bibinfo{pages}{441} (\bibinfo{year}{2008}).

\bibitem[{\citenamefont{Brunner et~al.}(2009)\citenamefont{Brunner, Gerardot,
  Dalgarno, Wust, Karrai, Stoltz, Petroff, and Warburton}}]{BrunnerScience09}
\bibinfo{author}{\bibfnamefont{D.}~\bibnamefont{Brunner}},
  \bibinfo{author}{\bibfnamefont{B.~D.} \bibnamefont{Gerardot}},
  \bibinfo{author}{\bibfnamefont{P.~A.} \bibnamefont{Dalgarno}},
  \bibinfo{author}{\bibfnamefont{G.}~\bibnamefont{Wust}},
  \bibinfo{author}{\bibfnamefont{K.}~\bibnamefont{Karrai}},
  \bibinfo{author}{\bibfnamefont{N.~G.} \bibnamefont{Stoltz}},
  \bibinfo{author}{\bibfnamefont{P.~M.} \bibnamefont{Petroff}},
  \bibnamefont{and} \bibinfo{author}{\bibfnamefont{R.~J.}
  \bibnamefont{Warburton}}, \bibinfo{journal}{Science}
  \textbf{\bibinfo{volume}{325}}, \bibinfo{pages}{70} (\bibinfo{year}{2009}).

\bibitem[{\citenamefont{Economou and Reinecke}(2007)}]{EconomouPRL07}
\bibinfo{author}{\bibfnamefont{S.~E.} \bibnamefont{Economou}} \bibnamefont{and}
  \bibinfo{author}{\bibfnamefont{T.~L.} \bibnamefont{Reinecke}},
  \bibinfo{journal}{Physical Review Letters} \textbf{\bibinfo{volume}{99}},
  \bibinfo{pages}{217401} (\bibinfo{year}{2007}).

\bibitem[{\citenamefont{Andlauer and Vogl}(2009)}]{AndlauerPRB09}
\bibinfo{author}{\bibfnamefont{T.}~\bibnamefont{Andlauer}} \bibnamefont{and}
  \bibinfo{author}{\bibfnamefont{P.}~\bibnamefont{Vogl}},
  \bibinfo{journal}{Physical Review B} \textbf{\bibinfo{volume}{79}},
  \bibinfo{pages}{045307} (\bibinfo{year}{2009}).

\bibitem[{\citenamefont{Doty and Gammon}(2009)}]{DotyPhys09}
\bibinfo{author}{\bibfnamefont{M.~F.} \bibnamefont{Doty}} \bibnamefont{and}
  \bibinfo{author}{\bibfnamefont{D.}~\bibnamefont{Gammon}},
  \bibinfo{journal}{Physics} \textbf{\bibinfo{volume}{2}}, \bibinfo{pages}{16}
  (\bibinfo{year}{2009}).

\bibitem[{\citenamefont{Kim et~al.}(2008)\citenamefont{Kim, Economou, Badescu,
  Scheibner, Bracker, Bashkansky, Reinecke, and Gammon}}]{KimPRL08}
\bibinfo{author}{\bibfnamefont{D.}~\bibnamefont{Kim}},
  \bibinfo{author}{\bibfnamefont{S.~E.} \bibnamefont{Economou}},
  \bibinfo{author}{\bibfnamefont{S.~C.} \bibnamefont{Badescu}},
  \bibinfo{author}{\bibfnamefont{M.}~\bibnamefont{Scheibner}},
  \bibinfo{author}{\bibfnamefont{A.~S.} \bibnamefont{Bracker}},
  \bibinfo{author}{\bibfnamefont{M.}~\bibnamefont{Bashkansky}},
  \bibinfo{author}{\bibfnamefont{T.~L.} \bibnamefont{Reinecke}},
  \bibnamefont{and} \bibinfo{author}{\bibfnamefont{D.}~\bibnamefont{Gammon}},
  \bibinfo{journal}{Physical Review Letters} \textbf{\bibinfo{volume}{101}},
  \bibinfo{pages}{4} (\bibinfo{year}{2008}).

\bibitem[{\citenamefont{Bracker et~al.}(2006)\citenamefont{Bracker, Scheibner,
  Doty, Stinaff, Ponomarev, Kim, Whitman, Reinecke, and Gammon}}]{BrackerAPL06}
\bibinfo{author}{\bibfnamefont{A.~S.} \bibnamefont{Bracker}},
  \bibinfo{author}{\bibfnamefont{M.}~\bibnamefont{Scheibner}},
  \bibinfo{author}{\bibfnamefont{M.~F.} \bibnamefont{Doty}},
  \bibinfo{author}{\bibfnamefont{E.~A.} \bibnamefont{Stinaff}},
  \bibinfo{author}{\bibfnamefont{I.~V.} \bibnamefont{Ponomarev}},
  \bibinfo{author}{\bibfnamefont{J.~C.} \bibnamefont{Kim}},
  \bibinfo{author}{\bibfnamefont{L.~J.} \bibnamefont{Whitman}},
  \bibinfo{author}{\bibfnamefont{T.~L.} \bibnamefont{Reinecke}},
  \bibnamefont{and} \bibinfo{author}{\bibfnamefont{D.}~\bibnamefont{Gammon}},
  \bibinfo{journal}{Applied Physics Letters} \textbf{\bibinfo{volume}{89}},
  \bibinfo{pages}{233110} (\bibinfo{year}{2006}).

\bibitem[{\citenamefont{Krenner et~al.}(2005)\citenamefont{Krenner, Sabathil,
  Clark, Kress, Schuh, Bichler, Abstreiter, and Finley}}]{KrennerPRL05}
\bibinfo{author}{\bibfnamefont{H.~J.} \bibnamefont{Krenner}},
  \bibinfo{author}{\bibfnamefont{M.}~\bibnamefont{Sabathil}},
  \bibinfo{author}{\bibfnamefont{E.~C.} \bibnamefont{Clark}},
  \bibinfo{author}{\bibfnamefont{A.}~\bibnamefont{Kress}},
  \bibinfo{author}{\bibfnamefont{D.}~\bibnamefont{Schuh}},
  \bibinfo{author}{\bibfnamefont{M.}~\bibnamefont{Bichler}},
  \bibinfo{author}{\bibfnamefont{G.}~\bibnamefont{Abstreiter}},
  \bibnamefont{and} \bibinfo{author}{\bibfnamefont{J.~J.}
  \bibnamefont{Finley}}, \bibinfo{journal}{Physical Review Letters}
  \textbf{\bibinfo{volume}{94}}, \bibinfo{pages}{057402}
  (\bibinfo{year}{2005}).

\bibitem[{\citenamefont{Gammon et~al.}(1996)\citenamefont{Gammon, Snow,
  Shanabrook, Katzer, and Park}}]{GammonPRL96}
\bibinfo{author}{\bibfnamefont{D.}~\bibnamefont{Gammon}},
  \bibinfo{author}{\bibfnamefont{E.~S.} \bibnamefont{Snow}},
  \bibinfo{author}{\bibfnamefont{B.~V.} \bibnamefont{Shanabrook}},
  \bibinfo{author}{\bibfnamefont{D.~S.} \bibnamefont{Katzer}},
  \bibnamefont{and} \bibinfo{author}{\bibfnamefont{D.}~\bibnamefont{Park}},
  \bibinfo{journal}{Physical Review Letters} \textbf{\bibinfo{volume}{76}},
  \bibinfo{pages}{3005} (\bibinfo{year}{1996}).

\bibitem[{\citenamefont{Stinaff et~al.}(2006)\citenamefont{Stinaff, Scheibner,
  Bracker, Ponomarev, Korenev, Ware, Doty, Reinecke, and
  Gammon}}]{StinaffScience06}
\bibinfo{author}{\bibfnamefont{E.~A.} \bibnamefont{Stinaff}},
  \bibinfo{author}{\bibfnamefont{M.}~\bibnamefont{Scheibner}},
  \bibinfo{author}{\bibfnamefont{A.~S.} \bibnamefont{Bracker}},
  \bibinfo{author}{\bibfnamefont{I.~V.} \bibnamefont{Ponomarev}},
  \bibinfo{author}{\bibfnamefont{V.~L.} \bibnamefont{Korenev}},
  \bibinfo{author}{\bibfnamefont{M.~E.} \bibnamefont{Ware}},
  \bibinfo{author}{\bibfnamefont{M.~F.} \bibnamefont{Doty}},
  \bibinfo{author}{\bibfnamefont{T.~L.} \bibnamefont{Reinecke}},
  \bibnamefont{and} \bibinfo{author}{\bibfnamefont{D.}~\bibnamefont{Gammon}},
  \bibinfo{journal}{Science} \textbf{\bibinfo{volume}{311}},
  \bibinfo{pages}{636} (\bibinfo{year}{2006}).

\bibitem[{\citenamefont{Krenner et~al.}(2006)\citenamefont{Krenner, Clark,
  Nakaoka, Bichler, Scheurer, Abstreiter, and Finley}}]{KrennerPRL06}
\bibinfo{author}{\bibfnamefont{H.~J.} \bibnamefont{Krenner}},
  \bibinfo{author}{\bibfnamefont{E.~C.} \bibnamefont{Clark}},
  \bibinfo{author}{\bibfnamefont{T.}~\bibnamefont{Nakaoka}},
  \bibinfo{author}{\bibfnamefont{M.}~\bibnamefont{Bichler}},
  \bibinfo{author}{\bibfnamefont{C.}~\bibnamefont{Scheurer}},
  \bibinfo{author}{\bibfnamefont{G.}~\bibnamefont{Abstreiter}},
  \bibnamefont{and} \bibinfo{author}{\bibfnamefont{J.~J.}
  \bibnamefont{Finley}}, \bibinfo{journal}{Physical Review Letters}
  \textbf{\bibinfo{volume}{97}}, \bibinfo{pages}{076403}
  (\bibinfo{year}{2006}).

\bibitem[{\citenamefont{Scheibner
  et~al.}(2007{\natexlab{a}})\citenamefont{Scheibner, Doty, Ponomarev, Bracker,
  Stinaff, Korenev, Reinecke, and Gammon}}]{ScheibnerPRB07}
\bibinfo{author}{\bibfnamefont{M.}~\bibnamefont{Scheibner}},
  \bibinfo{author}{\bibfnamefont{M.~F.} \bibnamefont{Doty}},
  \bibinfo{author}{\bibfnamefont{I.~V.} \bibnamefont{Ponomarev}},
  \bibinfo{author}{\bibfnamefont{A.~S.} \bibnamefont{Bracker}},
  \bibinfo{author}{\bibfnamefont{E.~A.} \bibnamefont{Stinaff}},
  \bibinfo{author}{\bibfnamefont{V.~L.} \bibnamefont{Korenev}},
  \bibinfo{author}{\bibfnamefont{T.~L.} \bibnamefont{Reinecke}},
  \bibnamefont{and} \bibinfo{author}{\bibfnamefont{D.}~\bibnamefont{Gammon}},
  \bibinfo{journal}{Physical Review B} \textbf{\bibinfo{volume}{75}},
  \bibinfo{pages}{245318} (\bibinfo{year}{2007}{\natexlab{a}}).

\bibitem[{\citenamefont{Doty et~al.}(2008)\citenamefont{Doty, Scheibner,
  Bracker, and Gammon}}]{DotyPRB08}
\bibinfo{author}{\bibfnamefont{M.~F.} \bibnamefont{Doty}},
  \bibinfo{author}{\bibfnamefont{M.}~\bibnamefont{Scheibner}},
  \bibinfo{author}{\bibfnamefont{A.~S.} \bibnamefont{Bracker}},
  \bibnamefont{and} \bibinfo{author}{\bibfnamefont{D.}~\bibnamefont{Gammon}},
  \bibinfo{journal}{Physical Review B} \textbf{\bibinfo{volume}{78}},
  \bibinfo{pages}{115316} (\bibinfo{year}{2008}).

\bibitem[{\citenamefont{Doty et~al.}(2006)\citenamefont{Doty, Scheibner,
  Ponomarev, Stinaff, Bracker, Korenev, Reinecke, and Gammon}}]{DotyPRL06}
\bibinfo{author}{\bibfnamefont{M.~F.} \bibnamefont{Doty}},
  \bibinfo{author}{\bibfnamefont{M.}~\bibnamefont{Scheibner}},
  \bibinfo{author}{\bibfnamefont{I.~V.} \bibnamefont{Ponomarev}},
  \bibinfo{author}{\bibfnamefont{E.~A.} \bibnamefont{Stinaff}},
  \bibinfo{author}{\bibfnamefont{A.~S.} \bibnamefont{Bracker}},
  \bibinfo{author}{\bibfnamefont{V.~L.} \bibnamefont{Korenev}},
  \bibinfo{author}{\bibfnamefont{T.~L.} \bibnamefont{Reinecke}},
  \bibnamefont{and} \bibinfo{author}{\bibfnamefont{D.}~\bibnamefont{Gammon}},
  \bibinfo{journal}{Physical Review Letters} \textbf{\bibinfo{volume}{97}},
  \bibinfo{pages}{197202} (\bibinfo{year}{2006}).

\bibitem[{\citenamefont{Doty et~al.}(2009)\citenamefont{Doty, Climente,
  Korkusinski, Scheibner, Bracker, Hawrylak, and Gammon}}]{DotyPRL09}
\bibinfo{author}{\bibfnamefont{M.~F.} \bibnamefont{Doty}},
  \bibinfo{author}{\bibfnamefont{J.~I.} \bibnamefont{Climente}},
  \bibinfo{author}{\bibfnamefont{M.}~\bibnamefont{Korkusinski}},
  \bibinfo{author}{\bibfnamefont{M.}~\bibnamefont{Scheibner}},
  \bibinfo{author}{\bibfnamefont{A.~S.} \bibnamefont{Bracker}},
  \bibinfo{author}{\bibfnamefont{P.}~\bibnamefont{Hawrylak}}, \bibnamefont{and}
  \bibinfo{author}{\bibfnamefont{D.}~\bibnamefont{Gammon}},
  \bibinfo{journal}{Physical Review Letters} \textbf{\bibinfo{volume}{102}},
  \bibinfo{pages}{047401} (\bibinfo{year}{2009}).

\bibitem[{\citenamefont{Climente et~al.}(2008)\citenamefont{Climente,
  Korkusinski, Goldoni, and Hawrylak}}]{ClimentePRB08}
\bibinfo{author}{\bibfnamefont{J.~I.} \bibnamefont{Climente}},
  \bibinfo{author}{\bibfnamefont{M.}~\bibnamefont{Korkusinski}},
  \bibinfo{author}{\bibfnamefont{G.}~\bibnamefont{Goldoni}}, \bibnamefont{and}
  \bibinfo{author}{\bibfnamefont{P.}~\bibnamefont{Hawrylak}},
  \bibinfo{journal}{Phys. Rev. B} \textbf{\bibinfo{volume}{78}},
  \bibinfo{pages}{115323} (\bibinfo{year}{2008}).

\bibitem[{\citenamefont{Scheibner
  et~al.}(2007{\natexlab{b}})\citenamefont{Scheibner, Ponomarev, Stinaff, Doty,
  Bracker, Hellberg, Reinecke, and Gammon}}]{ScheibnerPRL07}
\bibinfo{author}{\bibfnamefont{M.}~\bibnamefont{Scheibner}},
  \bibinfo{author}{\bibfnamefont{I.~V.} \bibnamefont{Ponomarev}},
  \bibinfo{author}{\bibfnamefont{E.~A.} \bibnamefont{Stinaff}},
  \bibinfo{author}{\bibfnamefont{M.~F.} \bibnamefont{Doty}},
  \bibinfo{author}{\bibfnamefont{A.~S.} \bibnamefont{Bracker}},
  \bibinfo{author}{\bibfnamefont{C.~S.} \bibnamefont{Hellberg}},
  \bibinfo{author}{\bibfnamefont{T.~L.} \bibnamefont{Reinecke}},
  \bibnamefont{and} \bibinfo{author}{\bibfnamefont{D.}~\bibnamefont{Gammon}},
  \bibinfo{journal}{Physical Review Letters} \textbf{\bibinfo{volume}{99}},
  \bibinfo{pages}{197402} (\bibinfo{year}{2007}{\natexlab{b}}).

\bibitem[{\citenamefont{Scheibner et~al.}(2008)\citenamefont{Scheibner, Yakes,
  Bracker, Ponomarev, Doty, Hellberg, Whitman, Reinecke, and
  Gammon}}]{ScheibnerNatPhys08}
\bibinfo{author}{\bibfnamefont{M.}~\bibnamefont{Scheibner}},
  \bibinfo{author}{\bibfnamefont{M.}~\bibnamefont{Yakes}},
  \bibinfo{author}{\bibfnamefont{A.~S.} \bibnamefont{Bracker}},
  \bibinfo{author}{\bibfnamefont{I.~V.} \bibnamefont{Ponomarev}},
  \bibinfo{author}{\bibfnamefont{M.~F.} \bibnamefont{Doty}},
  \bibinfo{author}{\bibfnamefont{C.~S.} \bibnamefont{Hellberg}},
  \bibinfo{author}{\bibfnamefont{L.~J.} \bibnamefont{Whitman}},
  \bibinfo{author}{\bibfnamefont{T.~L.} \bibnamefont{Reinecke}},
  \bibnamefont{and} \bibinfo{author}{\bibfnamefont{D.}~\bibnamefont{Gammon}},
  \bibinfo{journal}{Nat Phys} \textbf{\bibinfo{volume}{4}},
  \bibinfo{pages}{291} (\bibinfo{year}{2008}).

\bibitem[{\citenamefont{Bulaev and Loss}(2005)}]{BulaevPRL05}
\bibinfo{author}{\bibfnamefont{D.~V.} \bibnamefont{Bulaev}} \bibnamefont{and}
  \bibinfo{author}{\bibfnamefont{D.}~\bibnamefont{Loss}},
  \bibinfo{journal}{Physical Review Letters} \textbf{\bibinfo{volume}{95}},
  \bibinfo{pages}{076805} (\bibinfo{year}{2005}).

\bibitem[{\citenamefont{Luttinger and Kohn}(1955)}]{Luttinger55}
\bibinfo{author}{\bibfnamefont{J.~M.} \bibnamefont{Luttinger}}
  \bibnamefont{and} \bibinfo{author}{\bibfnamefont{W.}~\bibnamefont{Kohn}},
  \bibinfo{journal}{Physical Review} \textbf{\bibinfo{volume}{97}},
  \bibinfo{pages}{869} (\bibinfo{year}{1955}).

\bibitem[{\citenamefont{Rego et~al.}(1997)\citenamefont{Rego, Hawrylak, Brum,
  and Wojs}}]{RegoPRB97}
\bibinfo{author}{\bibfnamefont{L.~G.~C.} \bibnamefont{Rego}},
  \bibinfo{author}{\bibfnamefont{P.}~\bibnamefont{Hawrylak}},
  \bibinfo{author}{\bibfnamefont{J.~A.} \bibnamefont{Brum}}, \bibnamefont{and}
  \bibinfo{author}{\bibfnamefont{A.}~\bibnamefont{Wojs}},
  \bibinfo{journal}{Physical Review B} \textbf{\bibinfo{volume}{55}},
  \bibinfo{pages}{15694} (\bibinfo{year}{1997}).

\bibitem[{\citenamefont{Garcia et~al.}(1997)\citenamefont{Garcia,
  Medeiros-Ribeiro, Schmidt, Ngo, Feng, Lorke, Kotthaus, and
  Petroff}}]{GarciaAPL97}
\bibinfo{author}{\bibfnamefont{J.~M.} \bibnamefont{Garcia}},
  \bibinfo{author}{\bibfnamefont{G.}~\bibnamefont{Medeiros-Ribeiro}},
  \bibinfo{author}{\bibfnamefont{K.}~\bibnamefont{Schmidt}},
  \bibinfo{author}{\bibfnamefont{T.}~\bibnamefont{Ngo}},
  \bibinfo{author}{\bibfnamefont{J.~L.} \bibnamefont{Feng}},
  \bibinfo{author}{\bibfnamefont{A.}~\bibnamefont{Lorke}},
  \bibinfo{author}{\bibfnamefont{J.}~\bibnamefont{Kotthaus}}, \bibnamefont{and}
  \bibinfo{author}{\bibfnamefont{P.~M.} \bibnamefont{Petroff}},
  \bibinfo{journal}{Applied Physics Letters} \textbf{\bibinfo{volume}{71}},
  \bibinfo{pages}{2014} (\bibinfo{year}{1997}).

\bibitem[{\citenamefont{Ebbens et~al.}(2005)\citenamefont{Ebbens,
  Krizhanovskii, Tartakovskii, Pulizzi, Wright, Savelyev, Skolnick, and
  Hopkinson}}]{EbbensPRB05}
\bibinfo{author}{\bibfnamefont{A.}~\bibnamefont{Ebbens}},
  \bibinfo{author}{\bibfnamefont{D.~N.} \bibnamefont{Krizhanovskii}},
  \bibinfo{author}{\bibfnamefont{A.~I.} \bibnamefont{Tartakovskii}},
  \bibinfo{author}{\bibfnamefont{F.}~\bibnamefont{Pulizzi}},
  \bibinfo{author}{\bibfnamefont{T.}~\bibnamefont{Wright}},
  \bibinfo{author}{\bibfnamefont{A.~V.} \bibnamefont{Savelyev}},
  \bibinfo{author}{\bibfnamefont{M.~S.} \bibnamefont{Skolnick}},
  \bibnamefont{and}
  \bibinfo{author}{\bibfnamefont{M.}~\bibnamefont{Hopkinson}},
  \bibinfo{journal}{Physical Review B} \textbf{\bibinfo{volume}{72}},
  \bibinfo{pages}{4} (\bibinfo{year}{2005}).

\bibitem[{\citenamefont{Ferreira and Bastard}(1993)}]{FerreiraEPL93}
\bibinfo{author}{\bibfnamefont{R.}~\bibnamefont{Ferreira}} \bibnamefont{and}
  \bibinfo{author}{\bibfnamefont{G.}~\bibnamefont{Bastard}},
  \bibinfo{journal}{Europhysics Letters} \textbf{\bibinfo{volume}{23}},
  \bibinfo{pages}{439} (\bibinfo{year}{1993}).

\bibitem[{\citenamefont{Andreani et~al.}(1987)\citenamefont{Andreani,
  Pasquarello, and Bassani}}]{AndreaniPRB87}
\bibinfo{author}{\bibfnamefont{L.~C.} \bibnamefont{Andreani}},
  \bibinfo{author}{\bibfnamefont{A.}~\bibnamefont{Pasquarello}},
  \bibnamefont{and} \bibinfo{author}{\bibfnamefont{F.}~\bibnamefont{Bassani}},
  \bibinfo{journal}{Physical Review B} \textbf{\bibinfo{volume}{36}},
  \bibinfo{pages}{5887} (\bibinfo{year}{1987}).

\bibitem[{\citenamefont{Zutic et~al.}(2004)\citenamefont{Zutic, Fabian, and
  Das~Sarma}}]{ZuticRMP04}
\bibinfo{author}{\bibfnamefont{I.}~\bibnamefont{Zutic}},
  \bibinfo{author}{\bibfnamefont{J.}~\bibnamefont{Fabian}}, \bibnamefont{and}
  \bibinfo{author}{\bibfnamefont{S.}~\bibnamefont{Das~Sarma}},
  \bibinfo{journal}{Reviews of Modern Physics} \textbf{\bibinfo{volume}{76}},
  \bibinfo{pages}{323} (\bibinfo{year}{2004}).

\bibitem[{\citenamefont{Pfund et~al.}(2007)\citenamefont{Pfund, Shorubalko,
  Ensslin, and Leturcq}}]{PfundPRB07}
\bibinfo{author}{\bibfnamefont{A.}~\bibnamefont{Pfund}},
  \bibinfo{author}{\bibfnamefont{I.}~\bibnamefont{Shorubalko}},
  \bibinfo{author}{\bibfnamefont{K.}~\bibnamefont{Ensslin}}, \bibnamefont{and}
  \bibinfo{author}{\bibfnamefont{R.}~\bibnamefont{Leturcq}},
  \bibinfo{journal}{Physical Review B} \textbf{\bibinfo{volume}{76}},
  \bibinfo{pages}{161308} (\bibinfo{year}{2007}).

\bibitem[{\citenamefont{Economou and Reinecke}(2008)}]{EconomouPRB08}
\bibinfo{author}{\bibfnamefont{S.~E.} \bibnamefont{Economou}} \bibnamefont{and}
  \bibinfo{author}{\bibfnamefont{T.~L.} \bibnamefont{Reinecke}},
  \bibinfo{journal}{Physical Review B} \textbf{\bibinfo{volume}{78}},
  \bibinfo{pages}{5} (\bibinfo{year}{2008}).

\bibitem[{\citenamefont{Kato et~al.}(2003)\citenamefont{Kato, Myers, Driscoll,
  Gossard, Levy, and Awschalom}}]{KatoScience03}
\bibinfo{author}{\bibfnamefont{Y.}~\bibnamefont{Kato}},
  \bibinfo{author}{\bibfnamefont{R.~C.} \bibnamefont{Myers}},
  \bibinfo{author}{\bibfnamefont{D.~C.} \bibnamefont{Driscoll}},
  \bibinfo{author}{\bibfnamefont{A.~C.} \bibnamefont{Gossard}},
  \bibinfo{author}{\bibfnamefont{J.}~\bibnamefont{Levy}}, \bibnamefont{and}
  \bibinfo{author}{\bibfnamefont{D.~D.} \bibnamefont{Awschalom}},
  \bibinfo{journal}{Science} \textbf{\bibinfo{volume}{299}},
  \bibinfo{pages}{1201} (\bibinfo{year}{2003}).

\bibitem[{\citenamefont{Planelles et~al.}(2001)\citenamefont{Planelles,
  Jaskólski, and Aliaga}}]{PlanellesPRB01}
\bibinfo{author}{\bibfnamefont{J.}~\bibnamefont{Planelles}},
  \bibinfo{author}{\bibfnamefont{W.}~\bibnamefont{Jaskólski}},
  \bibnamefont{and} \bibinfo{author}{\bibfnamefont{J.~I.}
  \bibnamefont{Aliaga}}, \bibinfo{journal}{Physical Review B}
  \textbf{\bibinfo{volume}{65}}, \bibinfo{pages}{033306}
  (\bibinfo{year}{2001}).

\bibitem[{\citenamefont{Vurgaftman et~al.}(2001)\citenamefont{Vurgaftman,
  Meyer, and Ram-Mohan}}]{VurgaftmanJAP01}
\bibinfo{author}{\bibfnamefont{I.}~\bibnamefont{Vurgaftman}},
  \bibinfo{author}{\bibfnamefont{J.~R.} \bibnamefont{Meyer}}, \bibnamefont{and}
  \bibinfo{author}{\bibfnamefont{L.~R.} \bibnamefont{Ram-Mohan}},
  \bibinfo{journal}{Journal of Applied Physics} \textbf{\bibinfo{volume}{89}},
  \bibinfo{pages}{5815} (\bibinfo{year}{2001}).

\bibitem[{\citenamefont{Jacak et~al.}(1998)\citenamefont{Jacak, Hawrylak, and
  Wojs}}]{Jacak98}
\bibinfo{author}{\bibfnamefont{L.}~\bibnamefont{Jacak}},
  \bibinfo{author}{\bibfnamefont{P.}~\bibnamefont{Hawrylak}}, \bibnamefont{and}
  \bibinfo{author}{\bibfnamefont{A.}~\bibnamefont{Wojs}},
  \emph{\bibinfo{title}{Quantum dots}} (\bibinfo{publisher}{Springer Verlag,
  Berlin}, \bibinfo{year}{1998}).

\bibitem[{\citenamefont{Zhu et~al.}(2009)\citenamefont{Zhu, Karlsson,
  Byszewski, Rudra, Pelucchi, He, and Kapon}}]{ZhuSmall09}
\bibinfo{author}{\bibfnamefont{Q.}~\bibnamefont{Zhu}},
  \bibinfo{author}{\bibfnamefont{K.~F.} \bibnamefont{Karlsson}},
  \bibinfo{author}{\bibfnamefont{M.}~\bibnamefont{Byszewski}},
  \bibinfo{author}{\bibfnamefont{A.}~\bibnamefont{Rudra}},
  \bibinfo{author}{\bibfnamefont{E.}~\bibnamefont{Pelucchi}},
  \bibinfo{author}{\bibfnamefont{Z.~B.} \bibnamefont{He}}, \bibnamefont{and}
  \bibinfo{author}{\bibfnamefont{E.}~\bibnamefont{Kapon}},
  \bibinfo{journal}{Small} \textbf{\bibinfo{volume}{5}}, \bibinfo{pages}{329}
  (\bibinfo{year}{2009}).

\end{thebibliography}
\end{document}